\documentclass[reprint,amsmath,amssymb,aps,superscriptaddress]{revtex4-2}
\usepackage{graphicx}% Include figure files
\usepackage{dcolumn}% Align table columns on decimal point
\usepackage{bm}% bold math
\usepackage{hyperref}% add hypertext capabilities
\usepackage{braket}
\usepackage{enumitem}
\usepackage{xcolor}
\usepackage[normalem]{ulem}
\usepackage{transparent}
\usepackage{subcaption}
\usepackage[export]{adjustbox}
\usepackage{tikz}
\usetikzlibrary{positioning}
\graphicspath{{images/}}

\usepackage{pifont}

\newcommand{\vone}[1]{{}}
\newcommand{\vtwo}[1]{{#1}}

\begin{document}

% \preprint{APS/123-QED}

\title{Simple Fermionic backflow states via a systematically improvable tensor decomposition}

\author{Massimo Bortone}
    \email{massimo.bortone@kcl.ac.uk}
\affiliation{%
    Department of Physics, King’s College London, Strand, London WC2R 2LS, United Kingdom
}

\author{Yannic Rath}%
\affiliation{%
    Department of Physics, King’s College London, Strand, London WC2R 2LS, United Kingdom
}
\affiliation{National Physical Laboratory, Teddington, TW11 0LW, United Kingdom}

\author{George~H.~Booth}%
    \email{george.booth@kcl.ac.uk}
\affiliation{%
    Department of Physics, King’s College London, Strand, London WC2R 2LS, United Kingdom
}%

% \date{\today}

\begin{abstract}
    We present an effective ansatz for the wave function of correlated electrons that brings closer the fields of machine learning parameterizations and tensor rank decompositions.
    We consider a CANDECOMP/PARAFAC (CP) tensor factorization of a general backflow transformation in second quantization for a simple, compact and systematically improvable Fermionic state.
    This directly encodes $N$-body correlations without the ordering dependence of other tensor decompositions.
    We consider and explicitly demonstrate various controllable truncations, in the rank and range of the backflow correlations or magnitude of local energy contributions, in order to systematically affect scaling reductions to $\mathcal{O}[N^{3-4}]$.
    Benchmarking against small Fermi-Hubbard and chemical systems reveals an improvement over other NQS-like models, while extending towards larger strongly correlated {\em ab initio} systems demonstrates competitive accuracy with more established DMRG techniques on {\em ab initio} 2D hydrogenic lattices with realistic long-range Coulomb interactions.
\end{abstract}

\maketitle

%\tableofcontents

\section{Introduction}
Computational methods which can solve the physics of strongly correlated electrons play an important role in the study of molecular and condensed matter systems, where common perturbative or empirical density functional approaches fail.
In these systems, the interactions between electrons in some or all of their degrees of freedom contend with the kinetic energy of the electrons, leading to competition between localization and delocalization of the electronic structure, and the emergence of many remarkable properties and low-energy phases.
Electronic structure poses a particular challenge amongst the broader umbrella of quantum many-body problems~\cite{chooTwodimensionalFrustratedText2019} due to both charge and spin degrees of freedom, as well as the requirement for antisymmetry which significantly complicates the form of the solution.
However, these features are crucial in the understanding of the emergent physical behaviour in many technologically relevant advanced materials, from high-temperature superconductors to catalytic transition metal complexes.

A recent trend has emerged in the use of systematically improvable parameterized ans\"atze for quantum states, which hold the promise of an exact limit, providing confidence and the ability to internally validate results. 
Naturally, the universal approximators devised in the field of machine learning (ML) have inspired ans\"atze for this purpose, leading to the development of Neural Quantum States (NQS)~\cite{carleoSolvingQuantumManybody2017} with a wide range of network architectures, from restricted Boltzmann machines~\cite{nomuraRestrictedBoltzmannMachine2017,chooFermionicNeuralnetworkStates2020}, to networks with autoregressive~\cite{barrettAutoregressiveNeuralnetworkWavefunctions2022}, recurrent~\cite{hibat-allahRecurrentNeuralNetwork2020} or transformer architectures~\cite{spragueVariationalMonteCarlo2024} of differing depths and widths.
Concurrently, approaches based on kernel methods have also been considered~\cite{rathQuantumGaussianProcess2022,giulianiLearningGroundStates2023}.
All these parameterized ans\"atze can in principle approximate the complex functional dependencies between the probability amplitudes of the electronic configurations, and have proven capable of obtaining accurate results with minimal user intervention over a variety of systems.
Crucially, as long as the states can be sampled efficiently and the Hamiltonian of the system remains $k$-local, then the techniques of variational Monte Carlo (VMC) can stochastically optimize the parameters of the model to minimize the energy of the system, $E_{\theta} = \braket{\Psi_{\theta}|\hat{H}|\Psi_{\theta}}$, which then upper-bounds the true ground state energy, $E_0$.

It should be stressed that the use of systematically improvable ans\"atze in electronic structure is certainly not a new phenomenon with the emergence of NQS.
One of the most successful variational methods relies on tensor network states, which provide an improvable tensor factorization of the many-body amplitudes.
For a one-dimensional network, the efficient contraction of these amplitudes has led to the prominence of Matrix Product State (MPS) descriptions of correlated systems.
The single (hyper)parameter controlling the expressivity of the model is the `bond dimension', which can be quasi-continuously enlarged to describe higher levels of entanglement towards a complete model.
Importantly, the simple structure of these states also allows for new probes and insights into the emergent many-body physics of the model, and is able to characterize entanglement measures and structures~\cite{orusTensorNetworksComplex2019,eisertAreaLawsEntanglement2010a}.
Insights into the nature of the correlations and the entanglement can be harder to quantify for NQS, where the diversity of different architectures and model parameters can also cloud a clear path towards practical improvability for the states, while it can also be unclear how to precisely design optimal parameterizations.

In this work we return to tensor factorizations for a novel wave function parameterization, \vone{yet within the framework of NQS} \vtwo{inspired by the developments in the class of NQS descriptions of Fermionic quantum matter}.
There has been much research to indicate that the prodigious flexibility of complex NQS architectures is not being fully exploited for many correlated problems, due to the challenges in their optimization and initialization within the VMC framework~\cite{bukovLearningGroundState2021,inackNeuralAnnealingVisualization2022a}. 
Many simple parameterizations have performed as accurately as more complex forms, and a premium is placed on the compactness of the ansatz for ease of practical optimization alongside the overall flexibility.
The simpler form for these models can also potentially provide tools for easier interpretability and improvability of the many-body physics, and open avenues to alternative optimization strategies.
An example of this is the Gaussian process state (GPS), which was originally motivated via Bayesian regression as a systematically improvable kernel model with a single model parameter controlling the expressibility.
It was shown to achieve similar quality results to NQS, while often being more compact and open to novel insights~\cite{glielmoGaussianProcessStates2020,rathFrameworkEfficientInitio2023,giulianiLearningGroundStates2023}.
Further development exposed a duality of the GPS wave function model to an exponential of a CANDECOMP/PARAFAC (CP) tensor-rank decomposition of the wave function amplitudes in second quantization \cite{rathQuantumGaussianProcess2022}.
This represented an interesting simplification of the model, and suggests further developments \vone{to bring together the domains of NQS and} \vtwo{in the use of} tensor decompositions for systematically improvable descriptions of correlated states.
\vtwo{The potential for synergies between the two domains of variational state parameterizations in NQS and tensor decompositions has in fact already motivated the development of NQS architectures that incorporate MPS parameterizations~\cite{chenANTNBridgingAutoregressive2023,wuTensornetworkQuantumStates2023}.}

This is the topic of this work \vtwo{where we consider a simple and systematically improvable variational quantum state based on tensor factorization, with} \vone{in} application to general Fermionic systems which have proven a particular challenge for NQS methods to date.
Arguably the most success for {\em ab initio} electronic structure with NQS has come in a first quantized real-space formulation, where \vone{permutation-invariant multielectron orbitals are parameterized as a permutation-invariant neural network, before their product is explicitly antisymmetrized as a determinant (or short linear combination thereof)}%~\cite{pfauInitioSolutionManyelectron2020,hermannDeepneuralnetworkSolutionElectronic2020}.
\vtwo{the single-electron orbitals of a Slater determinant (or short linear combination thereof) have been replaced by multi-electron orbitals parametrized as permutation-invariant neural networks}~\cite{pfauInitioSolutionManyelectron2020,hermannDeepneuralnetworkSolutionElectronic2020,vonglehnSelfAttentionAnsatzAbinitio2022}.
In this work we consider a fixed basis and second quantization.
While this introduces a (necessarily incomplete) basis set approximation, it also allows for more flexibility in the choice of model (permutational invariance and antisymmetry are automatically enforced).
This also allows for problems to be defined in a finite and discrete space for the stochastic sampling where additional approximations can be devised and chemical insights from atomic orbital correlators are easily accessible.
Furthermore, this formulation allows for a straightforward treatment of core electrons and direct comparison to established quantum chemical methods, as well as natural application to multi-resolution and quantum embedding methodologies~\cite{nusspickelSystematicImprovabilityQuantum2022,sunQuantumEmbeddingTheories2016,kotliarElectronicStructureCalculations2006}.
The basis set approximation, in common with traditional quantum chemical methods, is also much studied with a number of approaches available which can substantially ameliorate it~\cite{delara-castellsCompleteBasisSet2001,gruneisPerspectiveExplicitlyCorrelated2017,boothExplicitlyCorrelatedApproach2012,yanaiCanonicalTranscorrelatedTheory2012}.

Direct application of NQS-like ans\"atze in second quantization have often struggled to clearly extend beyond state-of-the-art quantum chemistry, such as coupled-cluster methods (CCSD) or exact diagonalization (FCI), with results often restricted to small molecules and/or minimal basis sets~\cite{barrettAutoregressiveNeuralnetworkWavefunctions2022,chooFermionicNeuralnetworkStates2020,wuNNQSTransformerEfficientScalable2023,zhaoScalableNeuralQuantum2023}.
The commutation relations of second quantized operators enforced by a necessarily unphysical choice of ordering of the degrees of freedom can induce highly non-local and high-rank parity flips to the probability amplitudes.
In principle, these long-range structures can be described by NQS, but in practice are very difficult to model and to appropriately optimize within VMC frameworks, which have mainly been developed for quantum spin systems~\cite{chooFermionicNeuralnetworkStates2020}.
As such, finding better Fermion to spin (qubit) mappings to reduce the rank or range of these non-local parity changes is an active area of research~\cite{bravyiFermionicQuantumComputation2002,nysVariationalSolutionsFermiontoqubit2022,nysQuantumCircuitsSolving2023,derbyCompactFermionQubit2021}.
Alternatively, NQS-like states can be multiplied by an explicitly antisymmetric state (e.g. Slater determinant, Pfaffian or antisymmetrized geminal power) that will subsume much of the impact of these parity flips, at the cost of potentially limiting the rigorous systematic improvability of the resulting state.
Nevertheless, this approach in combination with symmetry-breaking and restoration has achieved impressive results in Fermionic models~\cite{nomuraRestrictedBoltzmannMachine2017,rathQuantumGaussianProcess2022,rathFrameworkEfficientInitio2023}.

Parallel to these developments, backflow transformations have been parameterized via neural networks as an alternative approach to describe Fermionic correlations in strongly interacting systems~\cite{tocchioRoleBackflowCorrelations2008,tocchioBackflowCorrelationsHubbard2011}.
This approach modifies the single-electron functions of a Slater determinant (or other antisymmetric function) to depend parametrically on many (potentially all $N$) electron coordinates in a configuration-dependent way. 
A highly related approach modifies the Slater determinant by coupling the physical degrees of freedom to a set of configuration-dependent auxiliary or `hidden' fermions, which can similarly be parameterized as a neural network~\cite{morenoFermionicWaveFunctions2022,liuUnifyingViewFermionic2024}.
The parameterization of these backflow-type states has undergone a similar development to other classes of variational wave functions, starting initially from physically-motivated few-body parameterizations~\cite{feynmanEnergySpectrumExcitations1956,wignerConstitutionMetallicSodium1934,kwonEffectsThreebodyBackflow1993,kwonEffectsBackflowCorrelation1998}, to a more general ML architecture which allows (in principle) for systematic improvability to exactness where each orbital and electron can arbitrarily change based on all other electronic positions.
These configuration-dependent orbitals in backflow states have been defined by a number of different ML architectures, both in real-space and discrete Fock space models, with and without an additional Jastrow factor in the parameterization~\cite{pfauInitioSolutionManyelectron2020,liuUnifyingViewFermionic2024,pesciaMessagePassingNeuralQuantum2024,romeroSpectroscopyTwodimensionalInteracting2024}.
They were shown to be effective in describing the ground states of Fermi-Hubbard models~\cite{luoBackflowTransformationsNeural2019,morenoFermionicWaveFunctions2022,zhouSolvingFermiHubbardtypeModels2024}, homogeneous electron gases~\cite{pesciaMessagePassingNeuralQuantum2024}, ultra-cold Fermi gases~\cite{kimNeuralnetworkQuantumStates2024} and (primarily in first quantization) {\em ab initio} molecular systems~\cite{pfauInitioSolutionManyelectron2020,vonglehnSelfAttentionAnsatzAbinitio2022,hermannDeepneuralnetworkSolutionElectronic2020,liuNeuralNetworkBackflow2024}, achieving energies comparable or surpassing those from Diffusion MC, as well as high accuracy coupled-cluster quantum chemical methods.

Here, we consider a particularly simple CP tensor rank decomposition for these configuration-dependent backflow orbitals, which allows for a straightforward yet systematically improvable form for the overall state~\cite{kiersStandardizedNotationTerminology2000,koldaTensorDecompositionsApplications2009}.
We show in Sec.~\ref{sec:methods} how this introduces explicit many-body correlated physics whilst retaining antisymmetry of the state.
Two previous approaches to combine tensor decompositions with backflow parameterizations have been considered in the literature; a backflow-inspired extension of the MPS ansatz~\cite{lamiMatrixProductStates2022} to build non-local entanglement beyond the native MPS tensor ordering constraints for spin systems, as well as a fixed tensor \vone{factorization} \vtwo{representation} of \vtwo{two-body} Fermionic backflow form~\cite{zhouSolvingFermiHubbardtypeModels2024}.
The latter study \vone{is more relevant to this work, but the particular choice of tensor factorization} did not \vtwo{factorize these backflow correlations and was constrained to a two-body form for these correlations.}\vone{admit a parameter by which the results could be systematically improved, instead relying on a Lanczos step to improve results.}
In contrast, the CP decomposition of the backflow parameterization introduced in this work overcomes these issues, providing a simple and improvable form \vtwo{for arbitrary rank correlations} which is invariant to orbital ordering.

We also develop a practical approach for {\em ab initio} systems to truncate the length scale of the backflow correlations, providing a further compression of the model with little loss of accuracy.
We apply this variational ansatz to find the ground state of (doped) Fermi-Hubbard models in Sec.~\ref{sec:fermi-hubbard}, significantly outperforming a comparable neural network backflow parameterization, as well as considering a challenging {\em ab initio} molecular electronic structure problem in Sec.~\ref{sec:water}.
In common with other studies, we find increasing the sampling of the VMC optimization important to improve results, indicating that despite the simple form of the state it is still challenging to optimize to the expressibility limit of the ansatz~\cite{chooFermionicNeuralnetworkStates2020,liuNeuralNetworkBackflow2024}.
%\mb{Are we happy with these two refs here?}
Nevertheless, we still achieve a comparable level of accuracy of more complex parameterizations.
In Sec.~\ref{sec:hydrogen}, we consider a $6 \times 6$ 2D lattice of hydrogen atoms as a step towards extended systems, with the CP\vtwo{D} backflow state comparing favourably to state-of-the-art DMRG calculations and significantly beyond the scope of exact approaches.
Finally, in Sec.~\ref{sec:scalability} we discuss the computational scaling of the method and approaches to reduce this as an outlook towards larger systems and widespread application.

\section{Backflow determinants via CP tensor-rank decomposition}\label{sec:methods}

The wave function for a system of $N$ interacting electrons can be defined in {\em first quantization} by assigning a unique label to each electron, and introducing real-space $\mathbf{r}_\alpha \in\mathbb{R}^3$ and spin $\sigma_\alpha \in\{\uparrow, \downarrow\}$ coordinates, so that $\Psi(\mathbf{x})=\Psi(\mathbf{x}_1,\dots,\mathbf{x}_N)$, with $\mathbf{x}_\alpha =(\mathbf{r}_\alpha,\sigma_\alpha)$.
The simplest wave function that satisfies the required antisymmetry is a Slater determinant of $N$ single-particle spin-orbitals, $\phi_i(\mathbf{x}_\alpha)$:
\begin{align}\label{eq:slater-determinant}
    \Phi_{0}(\mathbf{x}_1,\dots,\mathbf{x}_N) &= \frac{1}{\sqrt{N!}}
    \left\lvert
    \begin{matrix}
        \phi_{i}(\mathbf{x}_1) & \phi_{j}(\mathbf{x}_1) & \cdots & \phi_{k}(\mathbf{x}_1) \\
        \phi_{i}(\mathbf{x}_2) & \phi_{j}(\mathbf{x}_2) & \cdots & \phi_{k}(\mathbf{x}_2) \\
        \vdots & \vdots & \ddots & \vdots \\
        \phi_{i}(\mathbf{x}_N) & \phi_{j}(\mathbf{x}_N) & \cdots & \phi_{k}(\mathbf{x}_N)
    \end{matrix}
    \right\rvert, \\
    &= \mathcal{A}[\phi_i(\mathbf{x}_1)\phi_j(\mathbf{x}_2)\dots \phi_k(\mathbf{x}_N)],
\end{align}
where $\mathcal{A}$ antisymmetrizes and normalizes the subsequent product of orbitals with respect to exchange of their arguments.
We can consider these single-particle (molecular) orbitals as linear combinations of an underlying basis (e.g. {\em atomic orbitals}, AOs) $\chi_\mu(\mathbf{r})$, as
\begin{equation}
    \phi_{i}(\mathbf{r}) = \sum_{\mu=1}^L \varphi_{\mu i}\chi_\mu(\mathbf{r}),
\end{equation}
where $L$ is the size of this basis and $\varphi_{\mu i}$ are the coefficients of the linear combination.

The key-idea of backflow ans\"atze is to extend the Slater determinant by generalizing the single-particle orbitals to functions with non-linear parametric dependencies on all electron coordinates.
Historically this meant transforming the electron coordinates $\mathbf{r}_\alpha$ with a new set of coordinates $\mathbf{r}^{bf}_\alpha=\mathbf{r}_\alpha+\sum_{\beta\neq \alpha}\eta(|\mathbf{r}_\beta-\mathbf{r}_\alpha|)(\mathbf{r}_\beta-\mathbf{r}_\alpha)$, where the function $\eta(\mathbf{r})$ describes the effective displacement of the $\alpha$ electron due to the instantaneous position of the other electrons~\cite{lopezriosInhomogeneousBackflowTransformations2006}.
This configurational-dependence on all other electron positions can also be directly encoded into the variational parameters of a linear expansion of single-particle orbitals, as first introduced by Ref.~\cite{tocchioRoleBackflowCorrelations2008} for lattice models, yielding a new set of backflow orbitals $\phi^{bf}_{i}\left(\mathbf{r}_\alpha;\{\mathbf{r}_{/ \alpha}\}\right)$.
In an effort to improve the systematic description of these configuration-dependent backflow orbitals, recent work has proposed to model $\phi^{bf}_{i}\left(\mathbf{r}_\alpha;\{\mathbf{r}_{/ \alpha}\}\right)$ using neural networks~\cite{luoBackflowTransformationsNeural2019,pfauInitioSolutionManyelectron2020,vonglehnSelfAttentionAnsatzAbinitio2022}, ensuring that \vone{the electron labels in $\{\mathbf{r}_{/ \alpha}\}$}\vtwo{these functions} are invariant under permutation \vtwo{of the electron labels in $\{\mathbf{r}_{/ \alpha}\}$} to retain overall antisymmetry \vtwo{of the state}.

Within a {\em second quantization} representation, the permutational invariance of electrons and antisymmetry of the state is automatically ensured by the action and commutation relations of the second quantized operators, independent of the ansatz chosen.
A Slater determinant can thus be obtained from the vacuum state $\ket{0}$ by creating $N$ electrons in the corresponding single-particle orbitals as
\begin{equation}\label{eq:slater-determinant-2nd-quantization}
    \ket{\Phi_0} = \prod_{i=1}^N \hat{c}_{i}^{\dagger}\ket{0} = \prod_{i=1}^N \left(\sum_{\mu=1}^L\varphi_{\mu i}\hat{c}_\mu^{\dagger}\right)\ket{0},
\end{equation}
where $\hat{c}_\mu^{\dagger}$ ($\hat{c}_\mu$) is now the operator that creates (annihilates) an electron in the $\mu$-th basis state.
To model electron correlation, Eq.~\ref{eq:slater-determinant-2nd-quantization} can now be straightforwardly extended via analogy to the backflow transformations by including in each orbital a parametric dependence on the full instantaneous orbital occupation vector, $\mathbf{n}=(n_1,\dots,n_L)$, where $n_\mu$ indexes instantaneous occupancy of the four Fock states of spin-$\frac{1}{2}$ fermions in the chosen orthonormal representation of degree of freedom $\mu$.
This modifies the creation operator of orbital $i$ to be
\begin{equation}\label{eq:backflow-orbitals}
    \hat{c}^{\dagger}_{i} (\mathbf{n}) = \sum_{\mu=1}^{L}\varphi_{\mu i;\mathbf{n}}\hat{c}^{\dagger}_\mu ,
\end{equation}
resulting in an exact model, as each orbital can vary independently according to the instantaneous occupation over the full state.
However, it is of limited use as it is an over-parameterization of the full state, with an exponential number of variables.
We therefore consider a specific tensor-rank decomposition, the Canonical Decomposition (CANDECOMP) or Parallel Factor (PARAFAC) decomposition (CPD)\cite{kiersStandardizedNotationTerminology2000,koldaTensorDecompositionsApplications2009}.
This allows for a systematic and improvable decomposition of this tensor for each orbital into a polynomial and low-rank form that is independent of the choice of ordering of the degrees of freedom defining the occupation vector, $\mathbf{n}$.
The CP decomposition factorizes the occupation number vector over all states of Eq.~\ref{eq:backflow-orbitals} into a sum of $M$ tensor products, with each term in the product depending on each degree of freedom in the full occupation number vector, as
\begin{equation}\label{eq:tensor-rank-decomposition}
    \varphi^{\mathrm{CPD}}_{\mu i;\mathbf{n}} = \sum_{m=1}^M \prod_{\nu=1}^L \epsilon_{\mu i;n_{\nu} \nu m} .
\end{equation}
We now have a polynomially complex tensor of variational parameters for each orbital, $\epsilon_{\mu i;n_{\nu} \nu m}$, which encodes the correlation-driven modifications to orbital $i$ for the specific occupied degree of freedom $\mu$, based on the fact that state $\nu$ has a local occupation of $n_\nu$.
$M$ represents an improvable parameter describing the systematic coupling of the occupations across all possible occupation strings, providing an increasingly flexible description of higher-rank correlations in the state towards exactness.
We denote this single parameter controlling the flexibility of the model as its `support dimension', by analogy with the CP decomposition within Gaussian process states and kernel model definitions of quantum states~\cite{glielmoGaussianProcessStates2020,rathQuantumGaussianProcess2022,giulianiLearningGroundStates2023}.
This CP decomposition splits the $L$-dimensional indices indicating the $\mathbf{n}$-dependence of the orbital into a sum of products of rank-3 tensors, depending on each orbital and its occupation.
Since this is a simple product rather than matrix product, there is no change in the flexibility of these backflow orbitals with the ordering of the degrees of freedom, ensuring that there should be no explicit dependence on this choice (as found in tensor network states) or dimensionality of the system.

The proposed `CPD' backflow wave function is obtained by replacing the orbitals of the Slater determinant in Eq.~\ref{eq:slater-determinant-2nd-quantization} by those of Eq.~\ref{eq:tensor-rank-decomposition}, giving an explicitly antisymmetric state where all orbitals depend on the instantaneous occupation of all degrees of freedom,
\begin{equation}
    \ket{\Psi^\mathrm{CPD}} = \sum_{\mathbf{n}} \Psi^\mathrm{CPD}(\mathbf{n}) |\mathbf{n} \rangle ,
\end{equation}
    with
\begin{equation}\label{eq:cpd-state}
    \Psi^\mathrm{CPD}(\mathbf{n}) = \mathcal{A}[\varphi^{\mathrm{CPD}}_{\mu_1 i;\mathbf{n}} \varphi^{\mathrm{CPD}}_{\mu_2 j;\mathbf{n}} \dots \varphi^{\mathrm{CPD}}_{\mu_N k;\mathbf{n}}],
\end{equation}
where the antisymmetrizer acts with respect to the $N$ occupied orbitals of the configuration $\mathbf{n}$, given by $\mu_1, \mu_2 \dots \mu_N$.
This model can be evaluated naively via building a matrix and computing a determinant in $\mathcal{O}[N^2ML+N^3]$ cost, with each orbital evaluated according to Eq.~\ref{eq:tensor-rank-decomposition}.
However, for low-rank changes to $\mathbf{n}$ where only $\mathcal{O}[1]$ orbital occupations change, a fast updating scheme can be devised to reduce the scaling in the matrix build by a factor of $L$.
The update for each orbital in Eq.~\ref{eq:tensor-rank-decomposition} can be found in $\mathcal{O}[M]$ time by dividing out contributions from the previous occupations and multiplying by the new occupations, analogous to the approach in Ref.~\cite{rathFrameworkEfficientInitio2023}.
Since all configurational updates in VMC can be formulated in this way, the evaluation of configurational amplitudes of this CPD state can be reduced to $\mathcal{O}[N^2M + N^3]$.

The total number of variational parameters in this state (which in this work are all real) is therefore $\mathcal{O}[4 L^2 NM]$, where $L$ is the size of the underlying basis, $N$ is the number of electrons, and $M$ the `support dimension' controlling the flexibility of the model.
This scaling in terms of the evaluation of the model and number of parameters allows the standard techniques of VMC to be used for its sampling, optimization and extraction of observables.
We note that extending this CPD form to an explicit antisymmetrization of geminal two-particle states within a Pfaffian or antisymmetrized geminal power rather than single-particle orbitals of a determinant would also be possible in this framework and will be explored in the future~\cite{kimNeuralnetworkQuantumStates2024,louNeuralWaveFunctions2024}. %\mb{I've added these two refs which illustrate how real-space NQS have been extended with Pfaffians and AGPs}.

Unless otherwise indicated, in this work, we conserve a definite spin-polarization quantum number for each of the $N$ orbitals labelled $i, j, k, \dots$ in the product in Eq.~\ref{eq:cpd-state}, in which case the overall state must conserve $\hat{S}_z$ symmetry.
This is ensured by only allowing spin-orbital degrees of freedom with the same spin-polarization to be included in the expansion coefficients of the orbital (i.e. the $\mu$ labels in Eq.~\ref{eq:tensor-rank-decomposition}).
This allows the state for each $\mathbf{n}$ to factorize into a product of spin-up and spin-down determinants (which are allowed to independently optimize, analogous to an `unrestricted' single determinant).
An alternative (which is considered in Sec.~\ref{sec:fermi-hubbard}) is to form a `generalized' determinant by allowing spin-polarization to mix in each orbital definition, formally breaking $\hat{S}_z$ symmetry in the state.
This symmetry is nevertheless restored via the sampling of configurations with definite $\hat{S}_z$ in the Markov chain during the VMC procedure.
This $\hat{S}_z$ symmetry-breaking and projective restoration can improve results by allowing further flexibility in the state, but increases the cost in the evaluation of the determinant defining the amplitude by a factor of eight, and doubles the number of parameters (as $\mu$ labels spin-orbitals, not spatial orbitals).
Importantly, regardless of whether this spin symmetry is broken or not in the orbital definition, the backflow correlations act both for same-spin and opposite-spin correlations, with the orbital dependence in Eq.~\ref{eq:tensor-rank-decomposition} running over the spin-full occupations of all other degrees of freedom, $n_\nu$, ensuring that spin-dependent correlated physics is captured.

Finally, we note that, although the functional form of the configuration-dependent orbitals of Eq.~\ref{eq:tensor-rank-decomposition} is linear in $M$, it does not reduce to an uncorrelated determinant in the limit of $M=1$.
Correlated physics such as that captured via Gutzwiller or Jastrow correlators are included even in this limit, since the dependence between the instantaneous occupation of sites $\mu$ and $\nu$ can be independently addressed in a product form.
Indeed, full non-trivial $N$-body correlations are included even at $M=1$, as the exponentially large sum of products of these orbitals formed from the determinant in Eq.~\ref{eq:cpd-state} builds in an exponential sum of these $N$-fold products of variational parameters for each orbital.
This results in a highly-expressive state even for very low $M$, which is systematically improvable to exactness as $M$ is increased. \vtwo{We consider the expressibility of these states from an analytic perspective further below.}

\vtwo{
\subsection{Universality of the CPD backflow ansatz and Slater-Jastrow form}
}
\label{app:ansatz_universality}

\vtwo{The universality of the proposed CPD backflow ansatz directly stems from the universality of the CP decomposition.
This means that in the large-$M$ limit, the CP decomposition employed to model the orbitals $\varphi_{\mu i;\mathbf{n}}^\mathrm{CPD}$ according to Eq.~\eqref{eq:tensor-rank-decomposition}, can be chosen such that they are allowed to vary independently for each many-electron configuration $\mathbf{n}$.
This limit also implies a mapping between basis states (Slater determinants) and wavefunction amplitudes after anti-symmetrization which can represent {\em any} antisymmetric state without approximation error, according to the definition of the CPD backflow ansatz,
\begin{equation}
    \Psi^\mathrm{CPD}(\mathbf{n}) = \mathcal{A}[\varphi^{\mathrm{CPD}}_{\mu_1 i;\mathbf{n}} \varphi^{\mathrm{CPD}}_{\mu_2 j;\mathbf{n}} \dots \varphi^{\mathrm{CPD}}_{\mu_N k;\mathbf{n}}].
\end{equation}
This however {\em formally} requires $M$ to scale with the number of many-body configurations in the space (as expected for any exact parameterization), as we expand on below.}

\vtwo{To show this and make contact with other forms of parameterized states, we consider the subset of CPD states in which the backflow (many-electron) orbitals $\varphi_{\mu i;\mathbf{n}}^\mathrm{CPD}$ can factorize into a term which is independent of the specific configuration (an $\mathbf{n}$-independent `static' molecular orbital), and a term which is independent of the site index, but yet can depend on a CP decomposition of the specific many-electron configuration. This can be written as
\begin{equation}
    \varphi_{\mu i;\mathbf{n}}^\mathrm{factored-CPD} = \varphi_{\mu i} \times \left( \sum_{m=1}^M \prod_{\nu=1}^L \epsilon_{ i;n_{\nu} \nu m} \right)
\end{equation}
With this construction, a product of CP decompositions can be factored out of the determinant, bringing the backflow ansatz into the form of a Slater-Jastrow wavefunction, 
\begin{align}
    \label{eq:slater_jastrow_CPD}
    %\Psi^\textrm{CPD}(\mathbf{n}) = \prod_{i=1}^{N}  \left( \sum_{m=1}^M \prod_{\nu=1}^L \epsilon_{ i;n_{\nu} \nu m} \right) \times \mathcal{A}[\varphi_{\mu_1 i} \varphi_{\mu_2 j} \dots \varphi_{\mu_N k}].
    \Psi(\mathbf{n}) = &\left( \sum_{m=1}^M \prod_{\nu=1}^L \epsilon_{ i;n_{\nu} \nu m} \right) \left( \sum_{m=1}^M \prod_{\nu=1}^L \epsilon_{ j;n_{\nu} \nu m} \right) \cdots \nonumber \\ &\left( \sum_{m=1}^M \prod_{\nu=1}^L \epsilon_{ k;n_{\nu} \nu m} \right) \times \mathcal{A}[\varphi_{\mu_1 i} \varphi_{\mu_2 j} \dots \varphi_{\mu_N k}].
\end{align}
In this, a `configuration-universal' Slater determinant (defined by the $N$ orbitals labelled $i, j, \dots$) is multiplied by a product of $N$ CP decompositions, each of which depends on the local occupations of each site ($n_\nu$). This product of CP decompositions takes the place of the Jastrow factor.
Assuming that no two rows of the configuration-independent `static' Slater determinant orbital matrix, $\varphi_{\mu i}$, are linearly dependent such that the determinant always evaluates to a non-zero value~\cite{morenoFermionicWaveFunctions2022}, the universal approximator property of the CP decompositions in the prefactor allows this ansatz to define a one-to-one mapping from each many-electron configuration to an arbitrary wavefunction amplitude.}

\vtwo{This approach to factoring out a CPD decomposition from the Slater determinant shows that we formally require a support dimension $M$ which scales as the size of the Hilbert space and is therefore of little practical use.
Nonetheless, the representation according to Eq.~\eqref{eq:slater_jastrow_CPD} still provides insights into the ability of the ansatz to represent electronic quantum states of interest as a compact model with small support dimension, $M$.
From the consideration of the restricted version of the CPD state in this Slater-Jastrow form as shown in Eq.~\ref{eq:slater_jastrow_CPD}, we find that $M=1$ is sufficient to represent any single Slater determinant, as well as a site-dependent penalty function depending on its local occupation.
This encapsulates physically-relevant electronic correlation beyond the mean-field picture. As a specific example, we can consider a parameterization of a Gutzwiller factor of
\begin{equation}
\epsilon_{ i;n_{\nu} \nu m} = \begin{cases}
    e^{g_\nu} \quad \text{if } i = 1 \text{ and } n_\nu \equiv \uparrow \downarrow \\
    1 \quad \text{otherwise}\\
\end{cases},
\end{equation}
where $ n_\nu \equiv \uparrow \downarrow$ indicates a double occupancy of the $\nu^{\textrm{th}}$ site~\cite{gutzwiller1963effect, misawa2019mvmc}.
This modulates the Slater determinant with a factor depending on the double occupancy of the sites in each configuration, $\Psi(\mathbf{n}) \sim e^{\sum_\nu g_\nu n_{\nu,\uparrow} n_{\nu, \downarrow}} \times \mathcal{A}[\varphi_{\mu_1 i} \varphi_{\mu_2 j} \dots \varphi_{\mu_N k}]$, with parameters $g_\nu$. General forms for the $\epsilon_{ i;n_{\nu} \nu m}$ parameters even at $M=1$ will however also admit factorized non-local dependence on the site occupations beyond Gutzwiller form.}

\vtwo{In general, this simple factorization of the CPD state only represents a small subset of the parametrizations possible, which has a significantly larger variational flexibility even for $M=1$. This is because the factorization into a site-dependent term and Slater determinant does not need to be imposed, allowing a non-trivial coupling between these `orbital' and `site' effects at the level of the ansatz. This enlarges the span of states accessible within this decomposition at small $M$, and can therefore outperform `Slater-Jastrow' type factorizations where the Jastrow is taken to have a flexible form, such as those previously considered within the GPS family of states~\cite{rathFrameworkEfficientInitio2023}.}

\subsection{Initialization}\label{ssec:initialization}

A practical bottleneck in working with parameterized quantum states with many variational parameters can often be their reliable stochastic optimization.
This can be highly sensitive to the initialization of the state,\vone{ and t}\vtwo{ since random initialization of the parameters does not always guarantee a good overlap with the ground state, which can slow down or even prevent the optimization from converging to the true ground state.}
\vtwo{T}he simple functional form of the CPD backflow orbitals in Eq.~\ref{eq:tensor-rank-decomposition} allows for a straightforward and effective initialization of the variational parameters\vtwo{, without the requirement for pre-training~\cite{pfauInitioSolutionManyelectron2020,liuNeuralNetworkBackflow2024}}.
Specifically, the tensor $\epsilon_{\mu i;n_{\nu} \nu m}$ can be initialized to ensure that the CPD wave function exactly spans a given single determinant such as that found from a prior mean-field solution.
For most practical cases, this provides a good starting point for the VMC optimization.

In this work, we initialize from a restricted Hartree--Fock state, extracting the molecular orbital coefficients $\varphi^{\textrm{HF}}_{\mu i}$ in the basis in which the state is to be sampled.
The CPD variational parameter tensor can then be initialized as follows:
\begin{equation}\label{eq:cpd-init}
    \epsilon_{\mu i;n_{\nu} \nu m} = \mathcal{N}(0, \sigma) +
    \begin{cases}
        \varphi^{\mathrm{HF}}_{\mu i} & \text{if } m = 1 \text{ and } \nu = 1, \\
        1 & \text{if } m=1 \text{ and } \nu > 1 \\
        0 & \text{if } m>1 \text{ and } \nu \geq 1 \\
    \end{cases}
\end{equation}
where $\mathcal{N}(0, \sigma)$ is a random number drawn from a normal distribution with standard deviation $\sigma$.
This small amount of random noise is optional, but is added to the initialization in case the Hartree--Fock solution is too close to a local minimum of the optimization surface.

\subsection{Backflow truncation via exchange cutoff}\label{ssec:exchange-cutoff}

While the CPD backflow state only has a polynomial number of parameters, the $\mathcal{O}[4L^2 N M]$ scaling is still significantly higher than the native non-backflow (e.g. GPS) state, and there is are significant benefits in attempting to reduce this further with a controllable compromise on the flexibility of the state.
Largely redundant parameters in VMC add to statistical noise without improving accuracy and can be particularly deleterious in the optimization of the state~\cite{casulaCorrelatedGeminalWave2004,parkGeometryLearningNeural2020}. %\mb{I've added these two refs}.
In particular, the scaling with respect to the underlying basis size ($L$) is quadratic in the CPD state, and in this section we motivate a physical and black-box truncation of this scaling to further improve the overall performance of the state and enable access to larger systems.

We do this by restricting the number of degrees of freedom that the backflow parameterization considers for each $\mu$-indexed site, reducing it from $L$ to a new parameter $K$.
This can be motivated as a range-truncation of the backflow correlations, as has also been considered in other truncated expansions~\cite{tocchioBackflowCorrelationsHubbard2011}.
If application of this methodology was purely to local lattice models, then strictly truncating by a distance criteria would likely be sufficient to capture the dominant correlations.
However, we intend the methodology to be applied equally across lattice models and {\em ab initio} systems and therefore seek an alternative proxy to define the choice of entangled orbital subspace in which these backflow correlations are defined for each degree of freedom.
This is because an {\em ab initio} basis will necessarily be extended in space and perhaps not even able to be uniquely associated with an atomic centre.
Additionally, the inclusion of the long-range Coulomb interaction in these systems does not necessarily favour purely distance-based criteria.
We therefore take inspiration from {\em ab initio} formulations of the Density Matrix Renormalization Group (DMRG), where heuristics for the entanglement between two orbitals are necessary in order to find an approximately optimal ordering of the extended orbitals for an effective MPS ansatz.
While there are a number of options in the literature, it has been found that the importance of one orbital in describing the dominant correlations with another can be reasonably quantified by the magnitude of the exchange integral between them~\cite{olivares-amayaAbinitioDensityMatrix2015}, as
\begin{equation}
    \mathcal{K}_{\mu \nu} = \int\int d\mathbf{r}_1d\mathbf{r}_2\chi_{\mu}^*(\mathbf{r}_1)\chi_{\nu}^*(\mathbf{r}_1)\frac{1}{|\mathbf{r}_1-\mathbf{r}_2|}\chi_{\mu}(\mathbf{r}_2)\chi_{\nu}(\mathbf{r}_2) .
\end{equation}
This exchange-based metric should decay exponentially between localized orbitals, tending towards a flexible locality based truncation in the limit of fully local orbitals, while including the full range of the Coulomb interaction in the kernel.
More rigorous definitions of entanglement between orbitals such as their mutual information (pair entanglement entropy)~\cite{risslerMeasuringOrbitalInteraction2006} could also be used, but require an initial correlated level of theory on which to build these metrics.
Since we initialize the CPD backflow molecular orbitals from a Hartree-Fock calculation, the exchange matrix $\mathcal{K}_{\mu \nu}$ is readily available for no additional cost.
The set of $K$ most entangled orbitals for each orbital $\chi_\mu(\mathbf{r})$ according to this metric are selected, defining an $L \times K$ lookup table which maps to the relevant orbital indices $x_{\mu \nu} \in \{ 1, \dots, L\}$.
The choice of orbitals in the CPD decomposition of Eq.~\ref{eq:tensor-rank-decomposition} are therefore restricted as
\begin{equation}
    \varphi_{\mu i;\mathbf{n}} = \sum_{m=1}^M \prod_{\nu=1}^K \epsilon_{\mu i;n_{x_{\mu \nu}} \nu m} ,
\end{equation}
thus reducing the number of variational parameters to $\mathcal{O}(LKMN)$ and formally linear with the size of the system, assuming that $K$ is sufficiently large to capture the range of correlations around each degree of freedom.
As $K$ tends to $L$, the state returns to the original definition (albeit with an inconsequential reordering of sites in the backflow) giving the full flexibility of backflow correlations.

\begin{figure}
    \centering
    \begin{subfigure}[b]{0.45\linewidth}
        \includegraphics[width=\linewidth]{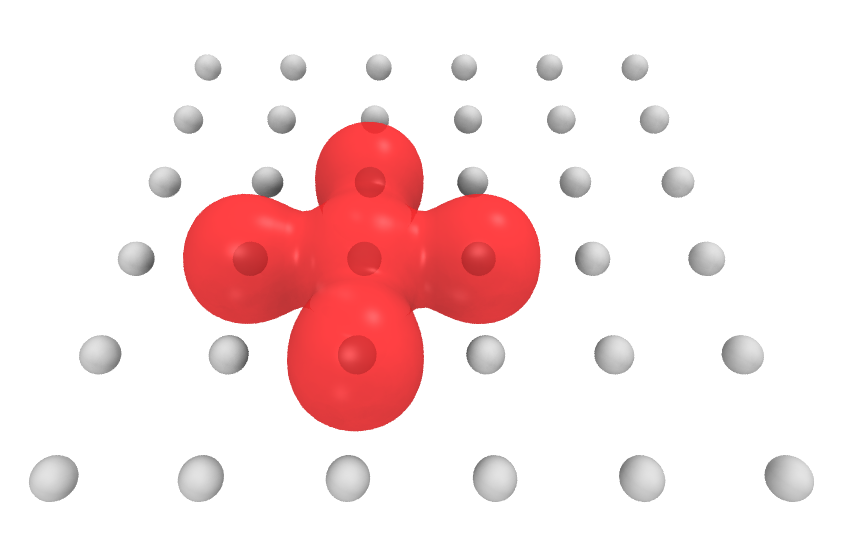}
        \caption{$d=1.0$ \AA}
    \end{subfigure}
    \hfill
    \begin{subfigure}[b]{0.45\linewidth}
        \includegraphics[width=\linewidth]{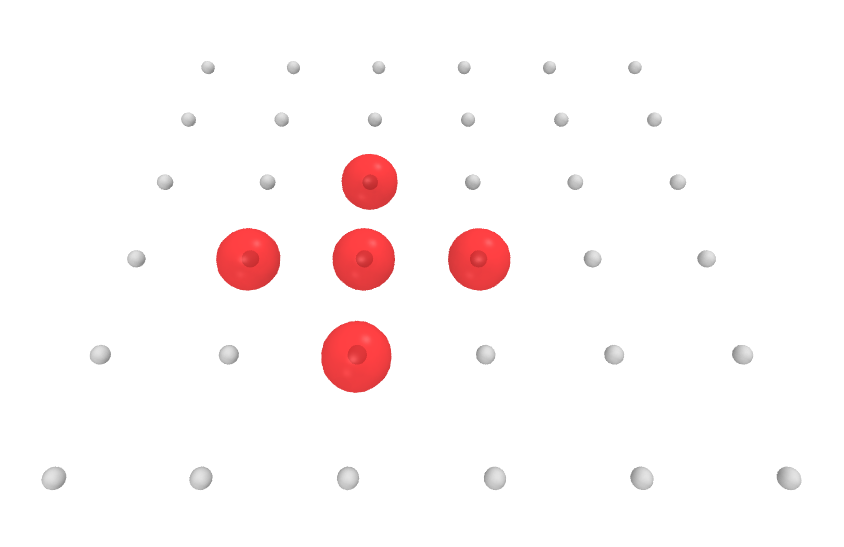}
        \caption{$d=2.0$ \AA}
    \end{subfigure}
    \caption{\vone{Electron density of the orbital subspace chosen via the exchange truncation in the range and rank of explicit backflow correlations.} 
    \vtwo{\textbf{Electron density of orbital subspaces selected via exchange truncation.}}
    \vone{We consider t}\vtwo{T}he $K=5$ \vtwo{orbital} subspace \vtwo{(red) is shown}\vone{ found} for a central atom in a $6\times 6$ lattice of hydrogen atoms in a STO-6G basis at both \vtwo{(a)} compressed and \vtwo{(b)} extended geometries.}
    \label{fig:hsheet-density}
\end{figure}

To illustrate the action of this exchange cutoff heuristic, in Figure~\ref{fig:hsheet-density} we consider the electron density of the $K=5$ most entangled orbitals about a specific atom for a $6 \times 6$ square grid of  {\em ab initio} hydrogen atoms in a Boys localized basis~\cite{fosterCanonicalConfigurationalInteraction1960} at two different interatomic distances, $d$. \vtwo{This truncation is used later for numerical results in Sec.~\ref{sec:hydrogen} to assess the accuracy of the truncation scheme. A choice of $K=5$ respects the local symmetries of each atom, as it enables each atom to be explicitly correlated via the backflow transformations with its four nearest neighbour atoms.}
\vone{As expected for a local representation,} \vtwo{As hoped, we find that the exchange cutoff protocol described automatically performs this} \vone{the exchange cutoff reduces to} selection of the nearest-neighbor atomic-localized orbitals around the chosen hydrogen atom in both geometries considered, providing a black-box metric to select the backflow subspace of correlations for each orbital via exploitation of locality of these correlations.
We note again that the product structure of the CPD ansatz will build longer-ranged and higher-rank correlations outside the chosen subset implicitly, albeit no longer explicitly for each orbital independently.

\section{Results} \label{sec:results}

In all results below we initialize the CPD backflow state from the restricted Hartree--Fock solution as outlined in Eq.~\ref{eq:cpd-init}, with a noise scale value $\sigma=0.01$.
We optimize parameters using the Stochastic Reconfiguration (SR) method~\cite{sorellaGeneralizedLanczosAlgorithm2001}, and when the number of parameters is larger than that of the samples, we take advantage of the recently introduced kernel formulation from Ref. ~\cite{rendeSimpleLinearAlgebra2024} to improve the computational cost of the optimization, as outlined further in Section~\ref{sec:scalability}.
On the Fermi-Hubbard model \vtwo{and the water molecule}, we found that a SR optimizer with RMSProp momentum regularization, as introduced in Ref.~\cite{lovatoHiddennucleonsNeuralnetworkQuantum2022}, outperforms standard SR, and we therefore use this optimizer for the results presented in Section~\ref{sec:fermi-hubbard} \vtwo{and~\ref{sec:water}}.
The final energies presented in the results are computed as averages over 50 independent energy evaluations with the final optimized parameters and a large sample size ($2^{16}$ for the Fermi-Hubbard model and the water molecule, and $2^{14}$ for the $6\times 6$ hydrogen lattice).
Error bars are computed as the standard deviation of these independent energy evaluations.

The VMC calculations are implemented in the {\tt NetKet} package~\cite{carleoNetKetMachineLearning2019, vicentiniNetKetMachineLearning2022}, which we interface with our own plugin module, {\tt GPSKet} for the required custom functionality.
For {\em ab initio} systems, Hartree--Fock orbital coefficients and Hamiltonians are supplied from {\tt PySCF}~\cite{sunPySCFPythonbasedSimulations2018,sunRecentDevelopmentsPySCF2020}.

\subsection{Fermi-Hubbard model}\label{sec:fermi-hubbard}

While the main ambition of this work is to apply the newly developed CPD backflow ansatz to {\em ab initio} systems, we first consider a small Fermi-Hubbard model on a 2D square lattice as a prototypical system for strongly correlated electrons, where comparison to exact results and neural-network parameterized backflow states from the literature are both available.
The Hamiltonian for this system is defined as
\begin{equation}
    \hat{H} = -t\sum_{\langle i,j\rangle,\sigma}\hat{c}_{i,\sigma}^{\dagger}\hat{c}_{j,\sigma} + U\sum_{i}\hat{n}_{i,\uparrow}\hat{n}_{i,\downarrow},
\end{equation}
where $\hat{c}_{i,\sigma}^{\dagger}$ ($\hat{c}_{i,\sigma}$) is the operator that creates (annihilates) a fermion with spin $\sigma$ on site $i$, $\hat{n}_{i,\sigma}=\hat{c}_{i,\sigma}^{\dagger}\hat{c}_{i,\sigma}$ is the number operator, $t$ is the hopping amplitude, and $U$ is the on-site interaction strength.
We apply the CPD backflow state, allowing for spin-polarization breaking and restoration of the orbitals as described in Sec.~\ref{sec:methods}, in the strong interaction regime at $U/t=8$ on a $4\times 4$ lattice with periodic boundary conditions, at half-filling ($n=N/L=1.0$) and in the hole doped case ($n=0.875$).
This hole-doped case is of particular interest as the point at which superconductivity and striped orders strongly compete and is much debated in the literature to date~\cite{zhengStripeOrderUnderdoped2017,sorellaSystematicallyImprovableMeanfield2023}.
We compare our results with those obtained by backflow ans\"atze based on neural networks (NNB) with similar numbers of parameters (${\sim}35,000$) taken from Ref.~\cite{luoBackflowTransformationsNeural2019} as well as exact diagonalization (ED)~\cite{dagottoStaticDynamicalProperties1992}.

\begin{figure}
    \centering
    \includegraphics[width=\linewidth]{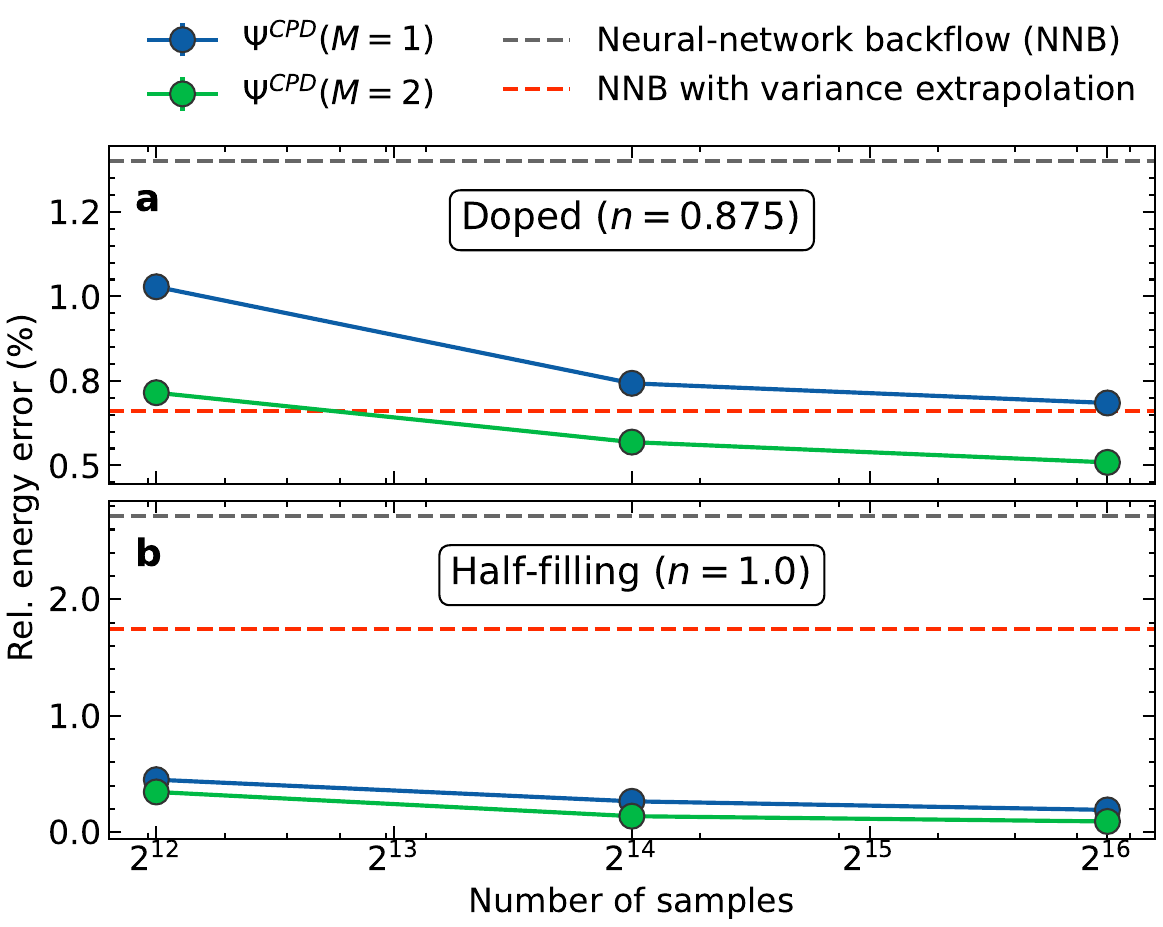}
    \caption{\vtwo{\textbf{Performance of the CPD and neural network backflow on the Fermi-Hubbard model.}}
    Percentage relative energy error for the ground state of the $4\times 4$ square Fermi-Hubbard model at $U/t=8$ compared to exact diagonalization results~\cite{dagottoStaticDynamicalProperties1992} at \vtwo{(\textbf{a})} a hole-doped filling of $n=0.875$ \vone{(upper panel)} and \vtwo{(\textbf{b})} half-filling \vone{(lower panel)}.
    CPD backflow results ($\Psi^{CPD}$) are show\vone{s}\vtwo{n} as a function of the configurational sample size in the optimization of the parameters, for two different model complexities of $M=1$ \vtwo{(blue)} and $M=2$ \vtwo{(green)}.
    Neural network backflow results \vtwo{(NNB, dashed lines)} are taken from Ref.~\cite{luoBackflowTransformationsNeural2019}.}
    \label{fig:hubbard2d-U8}
\end{figure}

In Fig.~\ref{fig:hubbard2d-U8} we show the percentage relative energy error compared to exact diagonalization for the CPD state of this system, plotted against the number of samples used in the Markov chain for each update of the parameters in the SR steps.
Our results significantly improve upon the comparable published neural-network backflow results for this system, even when these are extrapolated with respect to the complexity of the network architecture in the NNB ansatz.
We find percentage relative errors as low as $0.5\%$ for the doped case and $0.1\%$ for the half-filled case, which is competitive and within the scatter of other state-of-the-art techniques in the literature for this correlation regime \cite{simonscollaborationonthemany-electronproblemSolutionsTwoDimensionalHubbard2015}, albeit with this system too small to be compared in the thermodynamic limit. 

We also show the variational improvability as the support dimension $M$ of the CPD decomposition is increased from $M=1$ to $M=2$, with a systematic lowering of all energies found, leading to a maximum of $65,536$ parameters.
Nevertheless, we unfortunately find that it is still generally more advantageous to increase the number of configurational samples in the Markov chain than to formally increase the flexibility of the state by increasing $M$.
This is due to noise in the estimates of the expectation values required for the optimization of the CPD parameters, which amongst other things affects the inversion of the sampled quantum geometric tensor.
This indicates that we cannot be confident of a complete optimization to the global minimum of this state, despite the simple parameterization of the CPD form, with the optimization still limited more by noise in the samples than flexibility in the model, as found in many other studies of comparable states.
\vtwo{We will consider this behaviour more in the following section, but note that emerging optimization approaches, such as the SPRING algorithm~\cite{goldshlagerKaczmarzinspiredApproachAccelerate2024a}, will be able to be transferred to this setting and hold promise to boost the resulting performance of the CPD backflow state.}
\vtwo{However, despite these current limitations} we do find a reliable a systematic improvement in the optimized state as \vone{this}\vtwo{the} number of samples is increased, and a high level of accuracy overall for this correlated state.

\subsection{Water molecule}\label{sec:water}

We now consider {\em ab initio} molecular systems, which are described in second quantization by an electronic Hamiltonian of the form
\begin{equation}\label{eq:elec-hamiltonian}
    \hat{H} = \sum_{ij,\sigma} h^{(1)}_{ij}\hat{c}_{i,\sigma}^{\dagger}\hat{c}_{j,\sigma} + \frac{1}{2}\sum_{ijkl,\sigma \tau} h^{(2)}_{ijkl}\hat{c}_{i,\sigma}^{\dagger}\hat{c}_{j,\tau}^{\dagger}\hat{c}_{l,\tau}\hat{c}_{k,\sigma},
\end{equation}
where the sums run over the degrees of freedom in the system and $\sigma, \tau$ are binary spin variables.
The $h^{(1)}_{ij}$ matrix elements describe the kinetic energy operator and interaction with the external potential in these degrees of freedom, while the $h^{(2)}_{ijkl}$ terms model the Coulomb interaction between particles.
Compared to Fermi-Hubbard models, the computational complexity of these Hamiltonians is significantly increased by the $N^2(2L-N)^2$ scaling of the connected configurations required in evaluating the local energy (compared to $\mathcal{O}[N]$ terms in Hubbard and other lattice models).
Since the evaluation of the CPD wave function model at each configuration is $\mathcal{O}[N^2M+N^3]$ with the fast update (see Sec.~\ref{sec:methods}) this constrains the number of configurational samples that can be afforded.

For our initial benchmark system, we consider the water molecule in the 6-31G basis set at the equilibrium geometry used in Ref.~\cite{chooFermionicNeuralnetworkStates2020}.
While this seems an unassuming system from an electron correlation perspective, it has emerged as somewhat of a benchmark system in the Quantum Monte Carlo (QMC) community, where it has been studied extensively using a variety of ans\"atze~\cite{casulaCorrelatedGeminalWave2004,gurtubayDissociationEnergyWater2007,clarkComputingEnergyWater2011}.
Recently developed NQS architectures have struggled to reach state-of-the-art accuracy for this molecule, despite it still being of a size where exact diagonalization is possible.
Part of the issue with this comes from the fact that the weakly correlated physics and compact nature of the molecule mean that it is hard to define an appropriate representation for the basis which can enable efficient, faithful and representative sampling of the state with few configurational samples.

Minimizing the number of samples relies on finding a representation which can be faithfully approximated by a small stochastic selection of configurations, necessitating an orbital representation of the basis in which the wave function amplitudes are as flat as possible throughout the Hilbert space.
This maximizes the acceptance rates of the Metropolis-Hastings Markov chain growth, and ensures that as small a sample as possible can represent the wave function distribution.
Canonical bases of mean-field (e.g. Hartree--Fock) theories are therefore particularly poorly suited, as they are (away from very strong correlation) dominated by the configuration of a single Slater determinant.
These bases have been found for NQS with restricted Boltzmann machine architectures to obtain relatively large correlation energy errors ($\approx 5-10\%$) despite scaling up to $10^6$ configurational samples~\cite{chooFermionicNeuralnetworkStates2020}.
The development of autoregressive NQS models has been able to improve upon this by allowing a direct sampling algorithm of unique configurations that is not constrained by the limitations of the Metropolis-Hastings algorithm~\cite{barrettAutoregressiveNeuralnetworkWavefunctions2022}.
However, these models still require large sample sizes and have only been benchmarked in a STO-3G minimal basis set for this system~\cite{barrettAutoregressiveNeuralnetworkWavefunctions2022,zhaoScalableNeuralQuantum2023,wuNNQSTransformerEfficientScalable2023}.
Rather than changing the sampling algorithm, in a previous work, we considered the effect of different orbital representations for the configurational basis~\cite{rathFrameworkEfficientInitio2023}.
Following this, we consider orthogonal Foster-Boys orbitals for the configurations, localized over all degrees of freedom to minimize the physical spread of the resulting orbitals~\cite{fosterCanonicalConfigurationalInteraction1960}.

\begin{figure}
    \centering
    \includegraphics[width=\linewidth]{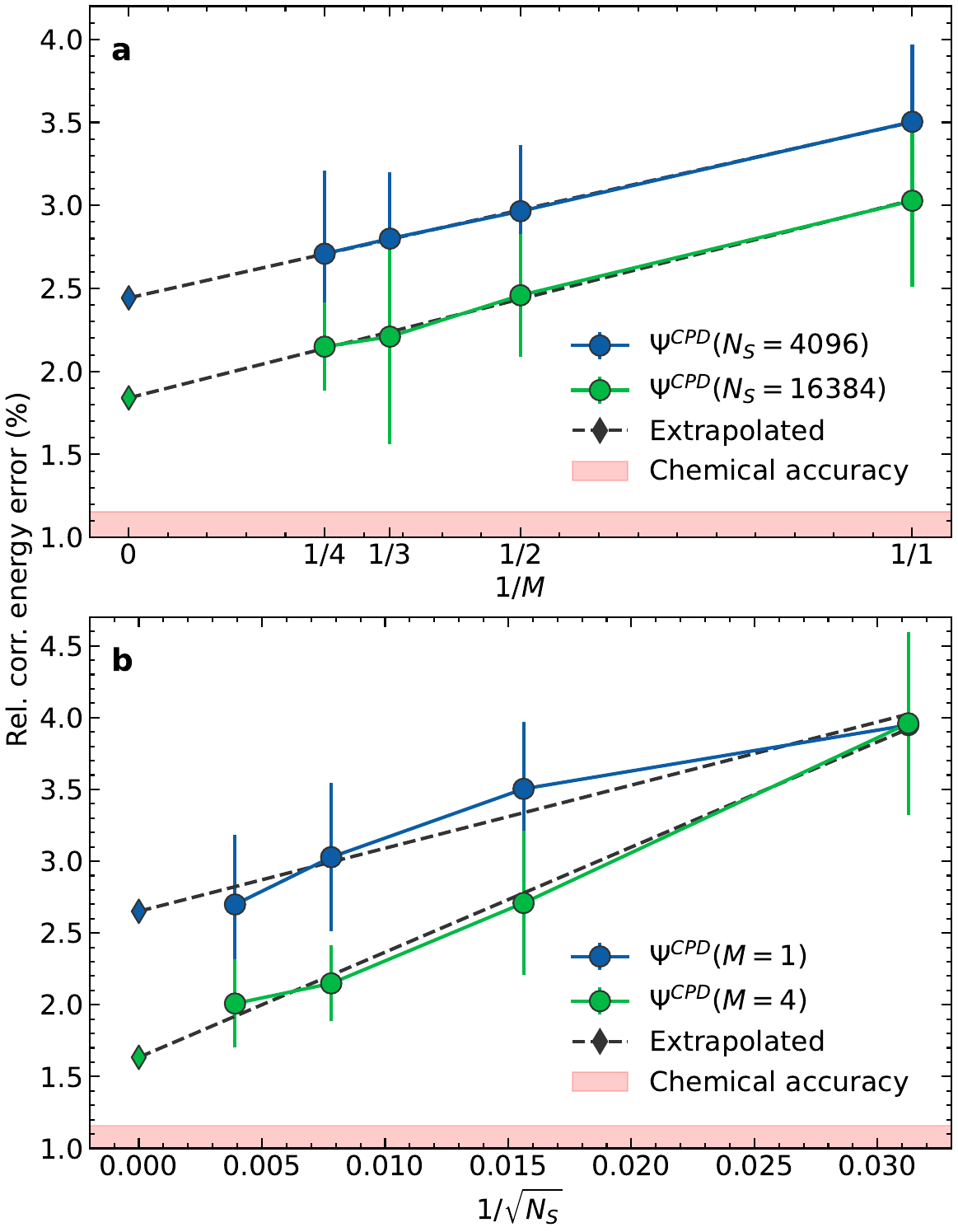}
    \caption{\vtwo{\textbf{Systematic improvability of the CPD backflow ansatz on the water molecule with respect to both model complexity ($M$) and stochastic samples ($N_S$).}
    Relative ground state correlation energy error (compared to exact diagonalization) for the CPD backflow ansatz ($\Psi^{CPD}$) on the water molecule (6-31G basis, equilibrium geometry as specified in Ref.~\cite{chooFermionicNeuralnetworkStates2020}) as a function of \textbf{(a)} support dimension $M$ and \textbf{(b)} number of configurational samples $N_S$. We also extrapolate these results to infinite $M$ or $N_S$ to provide the values in their infinite limit as shown by diamonds.}}
    %\textbf{a} The data for two CPD backflow models optimized with $N_S=4096$ (blue) and $N_S=16384$ (green) samples is extrapolated to the infinite support dimension limit (diamonds).
    %\textbf{b} The data for two CPD backflow models with $M=1$ (blue) and $M=4$ (green) is extrapolated to the infinite sample size limit (diamonds).}}
    \label{fig:h2o_syst_improv}
\end{figure}

\vtwo{The results in Fig.~\ref{fig:h2o_syst_improv}(a) show that the CPD backflow ansatz formulated in this local basis exhibits a strikingly systematic improvability, with the error decreasing inversely with the support dimension of the model, $M$.
We can use this empirical scaling to extrapolate the results to the infinite support dimension limit, which results in a relative correlation energy error of below $2\%$, for the ansatz optimized with $\mathcal{O}[10^4]$ configurational samples.
At infinite $M$ the model is complete, and the error therefore must arise from the incomplete optimization of the finite-$M$ models. We therefore also consider the improvability in $M$ for two different numbers of configurational samples, $N_S$ in the Markov chains used for each optimization step.
The $1/M$ decay of the error is clearly seen in both of these sample sizes, with the extrapolated model result decreasing towards exactness for increasing $N_S$.}

\vtwo{We analyse this trend more systematically in Fig.~\ref{fig:h2o_syst_improv}(b), where we show the convergence in $N_S$ for two different model complexity parameters $M$, showing a relatively robust $N_S^{-\frac{1}{2}}$ scaling in the error.}
%The improved efficiency of the local basis representation is further demonstrated in Fig.~\ref{fig:h2o_syst_improv}(b), where a systematic trend can be seen as the number of configurational samples $N_S$ in the Markov chain is increased.
\vtwo{This indicates that doubling the support dimension has a similar effect on reducing the error as quadrupling $N_S$.
This robustness and reliability in the error reduction is to be expected with increasing $M$, but is more surprising with increasing $N_S$.
It indicates that the noise introduced into the sampling at finite $N_S$ values is not simply changing the variance in the resulting energy, or indeed resulting in different optimized states due to convergence to different local minima in the landscape (where we would expect a wider scatter of optimized energies). Instead, the robustness and systematic trend in the results indicates that $N_S$ is controlling the intrinsic error of the optimization of the state in a more systematic fashion.
This could potentially arise from non-linear steps in the optimization protocol, and is something which requires further scrutiny going forwards.}

\begin{figure}
    \centering
    \includegraphics[width=\linewidth]{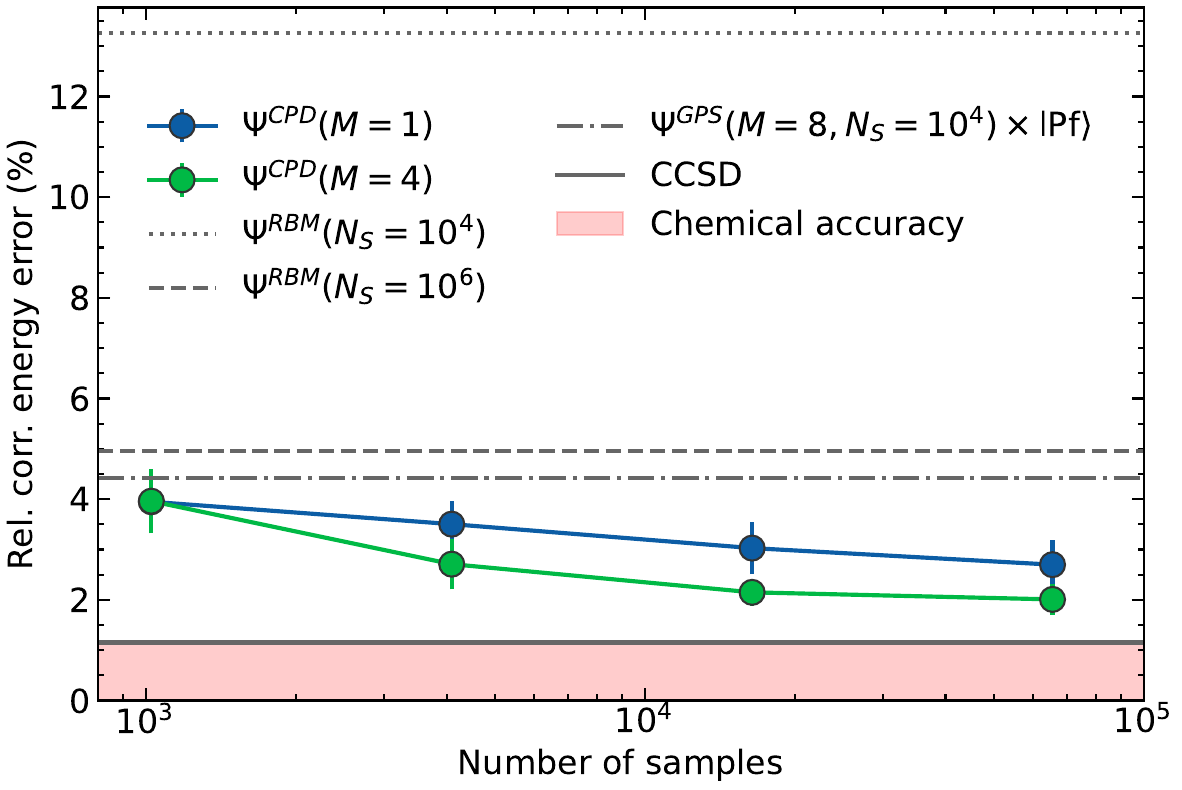}
    \caption{\vtwo{\textbf{Performance of the CPD backflow and other models on the water molecule.}}
    Relative ground state correlation energy error (compared to exact diagonalization) for CPD backflow ans\"atze ($\Psi^{CPD}$) on the water molecule (6-31G basis, equilibrium geometry as specified in Ref.~\cite{chooFermionicNeuralnetworkStates2020}) as the number of configurational samples \vtwo{$N_S$} in each Markov chain is increased.
    Two support dimensions corresponding to $M=1$ \vtwo{(blue)} and $M=4$ \vtwo{(green)} are shown.
    \vtwo{Comparison energies for the} Gaussian process state \vtwo{augmented by a symmetry-broken Pfaffian} (\vone{GPS}\vtwo{$\Psi^{GPS}\times\left|\text{Pf}\right\rangle$, dash-dotted line}) and restricted Boltzmann machine \vtwo{NQS} (\vone{RBM}\vtwo{$\Psi^{RBM}$, dotted and dashed lines}) \vone{comparison energies} are taken from Refs.~\cite{rathFrameworkEfficientInitio2023} and \cite{chooFermionicNeuralnetworkStates2020}, respectively, while the CCSD energy \vtwo{(solid line)} is calculated with PySCF~\cite{sunPySCFPythonbasedSimulations2018,sunRecentDevelopmentsPySCF2020}.}
    \label{fig:h2o_rel_corr_error}
\end{figure}

\vtwo{Comparing the accuracies obtained to previous state-of-the-art results in Figure \ref{fig:h2o_rel_corr_error}, we find that the CPD backflow state with $M=1$ ($6.7k$ parameters) already outperforms both the Gaussian process state augmented by a symmetry-broken Pfaffian ($585$ parameters)~\cite{rathFrameworkEfficientInitio2023} and the restricted Boltzmann machine NQS state ($728$ parameters)~\cite{chooFermionicNeuralnetworkStates2020} when optimized with $\mathcal{O}[10^3]$ configurational samples.
The accuracy is further improved when larger support dimensions and sample sizes are considered, with the CPD model at $M=4$ ($27k$ parameters) and $N_S\sim\mathcal{O}[10^4]$ outperforming the best NQS by $2\%$ in the relative correlation energy error.
While we still don't quite reach the level of accuracy of coupled cluster methods with singles and doubles (and the significant `chemical accuracy' hurdle -- albeit defined with respect to the finite basis set energy), to the best of our knowledge, these results represent the state-of-the-art for an NQS-like variational ansatz for this system.}

\vone{The results of the CPD backflow state for different number of configurational samples in this system are shown in Fig.~\ref{fig:h2o_rel_corr_error}.
Once again, we find that increasing the number of samples can affect the accuracy more significantly than increasing the complexity of the model, showing results going from $M=1$ ($6.7k$ parameters) to $M=4$ ($27k$ parameters).
Nevertheless, we outperform the accuracy of the native RBM state ($728$ parameters) with up to $10^6$ samples~\cite{chooFermionicNeuralnetworkStates2020}, as well as the GPS multiplied by a symmetry-broken Pfaffian state ($585$ parameters) with about $10^4$ samples~\cite{rathFrameworkEfficientInitio2023}, approaching the accuracy of coupled-cluster with singles and doubles level of theory which excels for weakly correlated equilibrium molecular systems such as this.
While we still don't quite reach this level (and the significant `chemical accuracy' hurdle -- albeit defined with respect to the finite basis set energy), to the best of our knowledge, these results represent the state-of-the-art for an NQS-like variational ansatz for this system, reaching a relative correlation energy error of $\approx 2\%$ with support dimension $M=4$ and using a comparatively small number of $\mathcal{O}[10^4]$ samples.}

\subsection{Towards hydrogen materials}\label{sec:hydrogen}

Extending the CPD backflow ansatz beyond benchmark studies and comparison to exact results, we consider a two-dimensional {\em ab initio} lattice of hydrogen atoms as a step towards combining strong correlation, long-range interactions and extended systems.
These hydrogenic systems have been studied by a variety of methods in the recent years given their simple specification and challenge of realistic interactions, whilst maintaining a close connection to the Fermi-Hubbard model~\cite{hachmannMultireferenceCorrelationLong2006,simonscollaborationonthemany-electronproblemSolutionManyElectronProblem2017,stellaStrongElectronicCorrelation2011,sinitskiyStrongCorrelationHydrogen2010,tsuchimochiStrongCorrelationsConstrainedpairing2009,rathFrameworkEfficientInitio2023,mottaGroundStatePropertiesHydrogen2020}.
In particular, different correlation regimes can be probed by simply changing the interatomic distance of the lattice, similar to tuning the interaction strength in the Fermi-Hubbard model.
However, crucially these hydrogen lattices require the accurate treatment of realistic long-range Coulomb interactions and their effects, which are not present in the Fermi-Hubbard model.
Accurately capturing the ground state of these systems for different interatomic distance is thus a challenging task for most quantum chemistry methods, as it requires a flexible and expressive model with a treatment of long-range interactions and high-energy scattering physics that gives rise to states of significantly different character.

\begin{figure}
    \centering
    \includegraphics[width=\linewidth]{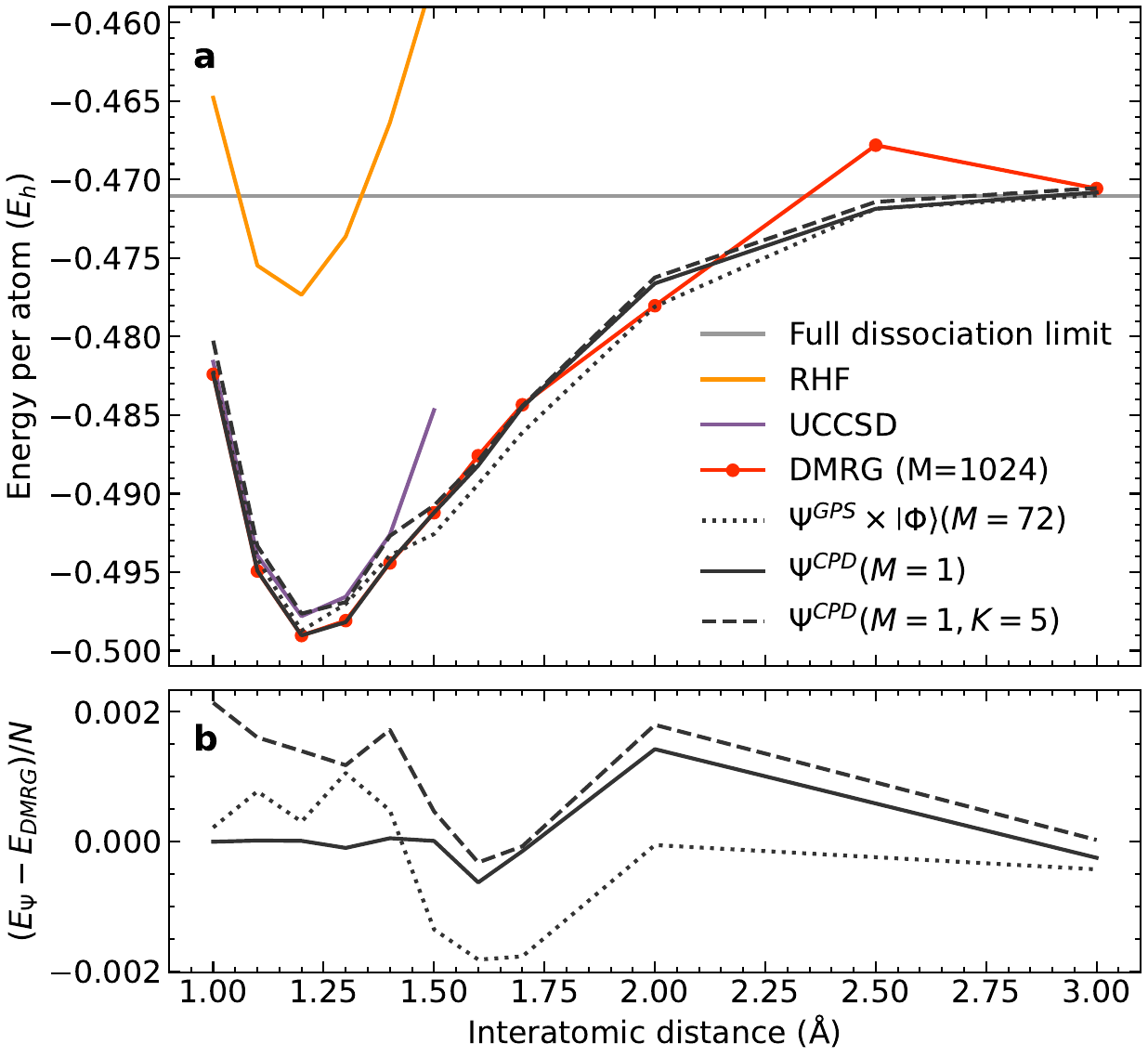}
    \caption{\vtwo{\textbf{Performance of the CPD backflow and other methods on the hydrogen lattice.}}
    \vtwo{\textbf{a}} Ground state energy per atom of a $6\times 6$ square hydrogen lattice in a STO-6G basis for increasing lattice constants.
    Shown \vone{(top)} are RHF (orange) and UCCSD energies (violet), while the horizontal grey line indicates the exact energy of the fully dissociated limit in this basis.
    The CPD backflow ansatz ($\Psi^{CPD}$) with support dimension $M=1$ (solid black line) is compared to a GPS ansatz augmented with a Slater determinant ($\Psi^{GPS} \times |\Phi \rangle$) with support dimension $M=72$~\cite{rathFrameworkEfficientInitio2023} (dotted black line), as well as energies obtained with DMRG (red dots and line) using a bond dimension of $M=1024$.
    A further CPD backflow ansatz with support dimension $M=1$ and a truncation in the backflow subspace to $K=5$ orbitals is shown as a dashed black line.
    The backflow CPD (GPS augmented by Slater determinant) VMC ans\"atze were optimized with $4096$ ($10,000$) samples, while DMRG results were obtained using the \texttt{block2} package~\cite{zhaiBlock2ComprehensiveOpen2023,sharmaSpinadaptedDensityMatrix2012}.
    \vtwo{\textbf{b}} \vone{The bottom panel shows e}\vtwo{E}nergy difference \vtwo{per atom} relative to DMRG of the CPD backflow ans\"atze and the GPS times Slater determinant ansatz.
    \vtwo{To aid the comparison, we have removed the outlier DMRG energy at $2.5$~\AA.}}
    \label{fig:hsheet-energy}
\end{figure}

In Figure \ref{fig:hsheet-energy}, we report the ground state energy per atom obtained from the CPD backflow ans\"atze (with and without the rank and range truncation of Sec.~\ref{ssec:exchange-cutoff} for the backflow transform to $K=5$ sites) for a $6\times 6$ hydrogen lattice in a minimal basis (STO-6G) \vtwo{with open boundary conditions}. \vtwo{The choice of $K=5$ for the range cutoff of the backflow ensures that the local symmetries of quantum fluctuations about each atomic site are preserved, as discussed in Sec.~\ref{ssec:exchange-cutoff}.}
We consider both compact (lower effective $U/t$) and extended (higher effective $U/t$) lattice structures by varying the interatomic distances all the way to essentially dissociated non-interacting hydrogen atoms.
We compare our results to energies obtained with restricted Hartree-Fock (RHF) and unrestricted coupled-cluster with single and double excitations (UCCSD), as well as an efficient {\em ab initio} implementation of density matrix renormalization group (DMRG) going up to bond dimension of $1024$ in a fully spin-adapted basis implemented in the \texttt{block2} package~\cite{zhaiBlock2ComprehensiveOpen2023,sharmaSpinadaptedDensityMatrix2012}.
For an additional comparison between contrasting approaches to Fermionic variational wave functions, we also include the results obtained from a GPS ansatz with support dimension $M=72$ acting as a Jastrow in front of a co-optimized Slater determinant~\cite{rathFrameworkEfficientInitio2023}, to compare the CPD backflow to this approach.
We optimize the CPD backflow and GPS multiplied by Slater determinant ans\"atze in a Boys localized basis for the orbitals, whereas for the DMRG results we rely on a split-localized basis, in which occupied and virtual orbitals are localized separately.
We found this choice to give the most consistent results for DMRG across the range of geometries studied.
\vtwo{Each optimization of the CPD backflow wave functions took $\approx 250$ GPU hours across 4 Nvidia A100 devices, whereas the DMRG runs took a total of $\approx 500$ CPU hours on an Intel(R) Core(TM) i9 device.}

The RHF and UCCSD description of this equation of state qualitatively breaks down quite early in this stretching coordinate, with UCCSD failing to converge beyond $1.5~\text{\AA}$.
Furthermore, the UCCSD exhibits quantitative error of $\sim 2~\text{mE}_h$ per atom even around equilibrium geometries, confirming that substantial correlation effects are present even in this regime. 
As another point of reference, the fully dissociated limit can be computed via exact diagonalization, where the assumption of simple energy extensivity from a single atom can be applied.
In this limit the energy is $\approx 0.03~\text{E}_h$ above the analytic result for the hydrogen atom due to the basis set incompleteness error, which nevertheless will exhibit a large degree of cancellation for energy differences along this changing geometry.
The DMRG provides the best variational comparison for this system, with (apart from $2.5~\text{\AA}$) the CPD and GPS results being within $2~\text{mE}_h$ per atom of this value.

The CPD backflow ansatz manages to quantitatively capture the features of the expected potential energy surface, reaching the correct dissociation limit at large interatomic distances, and showing an overall smooth transition from weak to strong correlation regimes.
We can directly compare different systematically improvable variational ans\"atze (DMRG, GPS multiplied by a Slater determinant and the CPD backflow), all of which are competitive and variationally optimal at different points in the changing physics of this system.
Around the equilibrium of this system, the CPD backflow and DMRG states are almost identical and variationally optimal amongst the comparison.
In the intermediate regime ($1.5 \leq d < 2.0~\text{\AA}$), the GPS ansatz augmented with a Slater determinant provides the best variational energies, despite (or perhaps because of) the smaller number of parameters ($\approx 12k$ vs. $\approx 187k$ for the CPD backflow ansatz without truncation and $\approx 26k$ for the one with).
An outlier appears to be the nearly dissociated limit of $2.5~\text{\AA}$ interatomic distance, where the DMRG energy appears erroneously high.
This could be due to a particularly large impact on the one-dimensional MPS topology used, the choice of basis for the orbitals or the DMRG sweep getting stuck in a local minimum.
Nevertheless, the other variational ansatz largely agree at this point.

Comparing the CPD backflow curves with and without the backflow truncation, we find (as expected) that the $K=5$ results are all variationally higher than the parent CPD backflow.
This truncation has a very small effect on the energies at larger interatomic distances, but becomes more significant around the equilibrium distance and mildly stretched geometries where it reaches a maximum error of $2~\text{mE}_h$ per atom.
This is expected as the range of the correlations in the compressed lattice will extend further than the stretched limit. 
Nonetheless, even with this restriction the ansatz is able to reach the coupled-cluster level of accuracy around equilibrium, and to outperform it on stretched geometries, with a reduction in the number of parameters compared to the parent model by more than a factor of seven.
This validates the exchange cutoff as a practical parameter reduction scheme for the CPD backflow ansatz, suggesting benefits in the study of larger systems, and potentially allowing for an increase in the support dimension of the model.

\begin{table}[h!]
    \centering
    \begin{tabular}{|l|c|c|c|c|}
    \hline
    Method & $D_e$ (eV)  & $r_e$ (\AA) & $w_e~(\text{cm}^{-1})$ & $w_ex_e (\text{cm}^{-1})$ \\
    \hline
    UCCSD                      & 0.163 & 1.225 & 2177.988 & 9.007 \\
    DMRG                       & 0.031 & 1.223 & 1900.038 & 36.409 \\
    $\Psi^{\textrm{CPD}}$                 & 0.029 & 1.219 & 1928.083 & 39.672 \\
    $\Psi^{\textrm{CPD}} (K=5)$           & 0.028 & 1.229 & 1865.462 & 38.586 \\
    $\Psi^{\textrm{GPS}}\times\ket{\Phi}$ & 0.029 & 1.237 & 1748.939 & 33.247 \\
    \hline
    \end{tabular}
    \caption{\vtwo{\textbf{Physical properties of the hydrogen lattice as obtained by different methods.}}
    Dissociation energy ($D_e$), equilibrium bond length ($r_e$), harmonic ($\omega_e$) and anharmonic ($\omega_e\chi_e$) frequencies in wavenumbers for the $6\times 6$ hydrogen lattice obtained by fitting a Morse potential to data from different quantum chemistry and variational methods described in this work.}
    \label{table:methods}
\end{table} 

To further compare the physical properties of the potential energy surface of this lattice as described by the different levels of theory, we fit a simple Morse potential at different interatomic distances ($r$), given by
\begin{equation}
    V(r) = D_e\left(1-e^{-a(r-r_e)}\right)^2+u ,
\end{equation}
where $D_e$ and $a$ control the depth and width of the well, $r_e$ is the equilibrium bond length, and $u$ is the energy offset.
Although the Morse potential is generally used for diatomic molecules, the symmetric stretching coordinate of this system is nevertheless well modelled by this form.
The differences in the quantum chemistry and VMC methods used to obtain the potential energy data are reflected in the variations of dissociation energy ($D_e$), equilibrium bond length ($r_e$), harmonic vibrational frequency ($\omega_e$), and anharmonicity constant ($\omega_e\chi_e$) presented in Table~\ref{table:methods}.

UCCSD is expected to accurately describe the correlated physics near equilibrium geometries, however the rapid divergence after this point renders even the harmonic vibrational frequencies unreliable.
In contrast, DMRG, which handles both the strong and weak correlations on a consistent level, presents a more accurate dissociation energy ($D_e = 0.031~\text{eV}$) and harmonic vibrational frequency ($\omega_e = 1900.038~\text{cm}^{-1}$), while yielding anharmonicity of the vibrational motion of the atomic lattice describing the beyond-parabolic nature of the binding as $\omega_e\chi_e = 36.409~\text{cm}^{-1}$.
The values obtained from CPD backflow ans\"atze with and without truncation closely track those from DMRG.
On the other hand, while agreeing on the dissociation energy, the GPS multiplied by Slater determinant ansatz stands out amongst the variational methods with a marginally softer bond, with a larger equilibrium lattice parameter ($r_e = 1.237~\text{\AA}$) and the lowest harmonic vibrational frequency ($\omega_e = 1748.939~\text{cm}^{-1}$).
Overall, the methods agree on an equilibrium bond length around $1.22~\text{\AA}$ and a dissociation energy of $0.03~\text{eV}$ (except UCCSD), with variations in the harmonic and anharmonic wavenumbers.

These variations highlight the strengths and limitations of each method in modelling the potential energy surfaces and vibrational properties of hydrogen materials.
Taking all these results into consideration, the CPD backflow ansatz emerges as a competitive method for the study of strong electron correlation, providing a variational description of the ground state of a two-dimensional lattice of hydrogen atoms that is in good agreement with other state-of-the-art methods, while being able to capture strong correlations and anharmonic effects in the system in a low-energy basis.

\subsubsection{Spin-spin correlations}

The local atomic basis framework of the CPD backflow ansatz allows for the straightforward computation of atom-resolved expectation values for further insights into the electronic structure.
In the context of hydrogen materials, local spin-spin correlation functions are of particular interest, as they can provide insights into the nature of the ground state of the system, and the emergence of magnetic order.
By analogy with Hubbard models, we would expect some anti-ferromagnetic order to emerge in the electronic structure of this system, with this order decaying algebraically in the thermodynamic limit.
However, in the presence of long-range interactions this behaviour is far from confirmed in two-dimensions.
While admittedly far from this thermodynamic limit, we consider the two-point spin-spin correlation function $C(r)$ between the centre of the $6\times 6$ hydrogen lattice, and atoms at a distance $r$ from the centre.
We can define this function via instantaneous (equal-time) spin-spin correlators $\left\langle \hat{S}^z_{\vec{r}_a}\hat{S}^z_{\vec{r}_b} \right\rangle$ between two atoms as
\begin{equation}
    C(r) = \frac{1}{N_{bulk}} \sum_{\vec{r}_a\in\text{bulk}}\sum_{|\vec{r}_a-\vec{r}_b|=r} \left\langle \hat{S}^z_{\vec{r}_a}\hat{S}^z_{\vec{r}_b} \right\rangle,
\end{equation}
where $\vec{r}_a$ and $\vec{r}_b$ are the positions of atoms $a$ and $b$, and $N_{bulk}$ is the number of equivalent atoms in the centre that we average over (four).
We use the atom-centered atomic orbitals themselves as natural projectors for the spin operators of each atom.
More details about the calculation of the instantaneous spin-spin correlation function can be found in Appendix~\ref{app:spin-spin-correlations}.

\begin{figure}
    \centering
    \includegraphics[width=\linewidth]{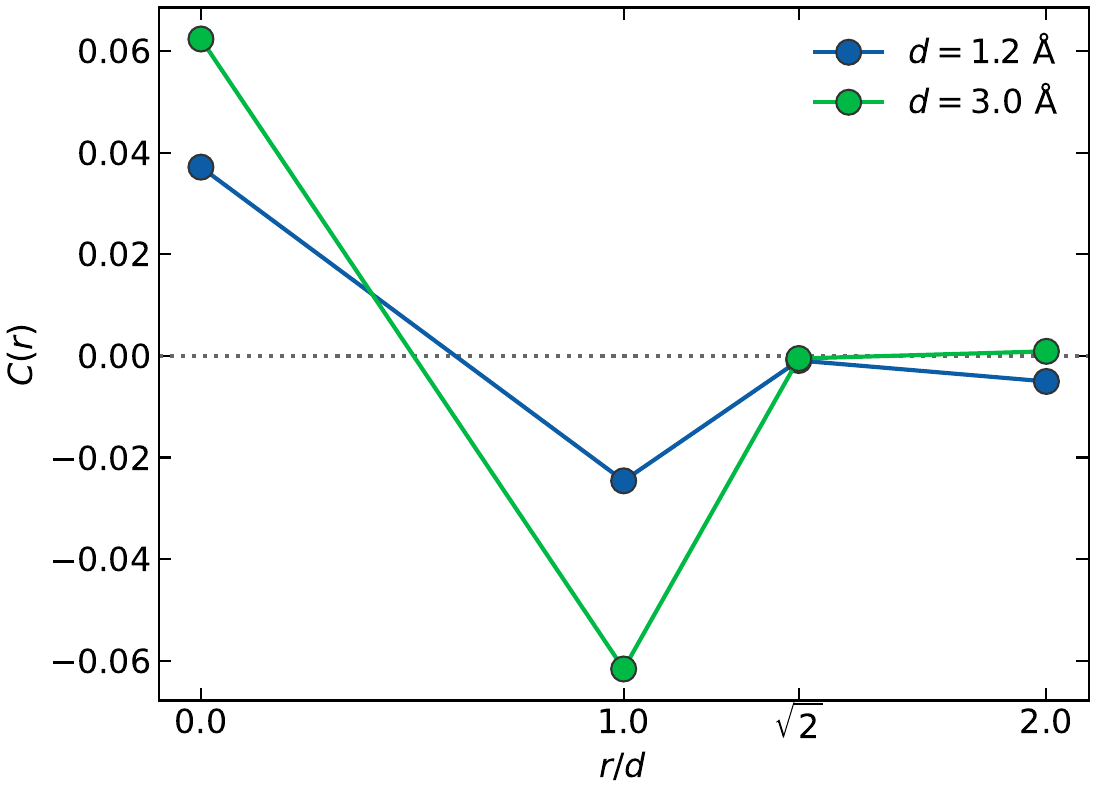}
    \caption{\vtwo{\textbf{Spin-spin correlations in the hydrogen lattice.}}
    Two-point instantaneous spin-spin correlation function for the ground state of the $6\times 6$ hydrogen lattice between atoms at two different interatomic distances $d=1.2$~\AA~\vtwo{(blue)} and $d=3.0$~\AA~\vtwo{(green)}. Correlation function is given as a function of normalized radial interatomic distance between atoms in the correlator, showing nearest, next-nearest and next-next-nearest magnetic correlations.
    The correlators were computed from the optimized CPD backflow ansatz with $M=1$.}
    \label{fig:hsheet-corr}
\end{figure}

In Figure \ref{fig:hsheet-corr}, we compare the radial spin-spin correlation function for the ground state approximation obtained with the CPD backflow ansatz at near-equilibrium interatomic distance $d=1.2$ \AA, and at a large stretching of $d=3.0$ \AA, normalized for the changing inter-atomic distances.
We find the emergence of the short-range anti-ferromagnetic order in the material, as anticipated by analogy with Hubbard models.
The magnitude of this antiferromagnetic order increases with increasing interatomic separation, again keeping with anticipated Hubbard behaviour of increasing $U/t$ values.
However, this order is very short ranged, with the spin in the extended lattice not directly affecting a lattice site beyond its nearest neighbours.
At more compressed geometries, this order does extend beyond this to the outer atoms in the lattice (next-next-nearest-neighbours), due to the shorter distance in real space, but the overall magnitude of these magnetic correlations is reduced.

\section{Scaling}\label{sec:scalability}

As illustrated in the results above, the CPD backflow ansatz performs well on small Fermionic systems, but further developments for scaling to significantly larger systems are still required for this to become a clearly competitive method for the wider electronic structure community.
Simplifying the scaling to assume a general growth of both the basis and electron number such that $N \sim L$ and assuming $M$ and $K$ are independent of system size, the parameters grow with system size as $N_P \sim \mathcal{O}[N^3]$ for the full ansatz, with the subspace truncation of the backflow reducing this asymptotically to $N_P \sim \mathcal{O}[N^2]$ (see Sec.~\ref{ssec:exchange-cutoff}).
The fast updating of backflow orbitals also enables the evaluation of the wave function log-amplitudes to be performed in $\mathcal{O}[MN^2 + N^3]$ (regardless of whether a backflow truncation is applied).
However, we find in practice the determinant evaluation has a significantly smaller prefactor than the construction of the orbitals, so that the dominant scaling is rather $\mathcal{O}[N^2]$ for small to medium-sized systems.
Given that the number of terms in general second quantized {\em ab initio} hamiltonians scales as $\mathcal{O}[N^4]$, the resulting scaling of the local energy evaluation is then $\mathcal{O}[N^7]$ for the CPD backflow state in the asymptotic limit, or $\mathcal{O}[N^6]$ for small to medium-sized systems.
While this should be competitive with accurate quantum chemical methods such as coupled-cluster, it is clear that the prefactor is significantly larger.

Rather than just the local energy evaluation, we should also consider the computational scaling for the update of the parameters.
For larger numbers of parameters (such as the $\sim187,000$ of the hydrogen lattice above), their update used to be the main bottleneck for VMC large-scale ans\"atze when using the original SR algorithm~\cite{sorellaGeneralizedLanczosAlgorithm2001}. 
For a model with $N_P$ parameters, SR would scale as $\mathcal{O}(N_P^3)$, since it involves inverting the $N_P \times N_P$ quantum geometric tensor matrix.
\vtwo{This is not the case for recently introduced alternative formulations of SR, such as minimum-step SR~\cite{chenEmpoweringDeepNeural2024}, the kernel formulation of SR~\cite{rendeSimpleLinearAlgebra2024} or SPRING~\cite{goldshlagerKaczmarzinspiredApproachAccelerate2024a}.}
\vone{This is not the case with}\vtwo{In particular for minimum-step SR and} the kernel formulation of SR \vone{recently introduced in Ref.~\cite{rendeSimpleLinearAlgebra2024}, which takes advantage of} \vone{a} simple linear algebra \vone{identity} \vtwo{identities were used} to reduce the dimension of the matrix that is inverted in the SR algorithm from $N_P$ tos $N_S$, i.e. the number of samples used during the optimization.
When $N_P\gg N_S$, as in large-scale models, the scaling of the parameter update becomes $\mathcal{O}(N_S^2N_P+N_S^3)$, i.e. linear in the number of parameters.
Thus, the evaluation of the local energy remains the computational bottleneck of the algorithm in the case of {\em ab initio} systems.

\begin{figure}
    \centering
    \includegraphics[width=\linewidth]{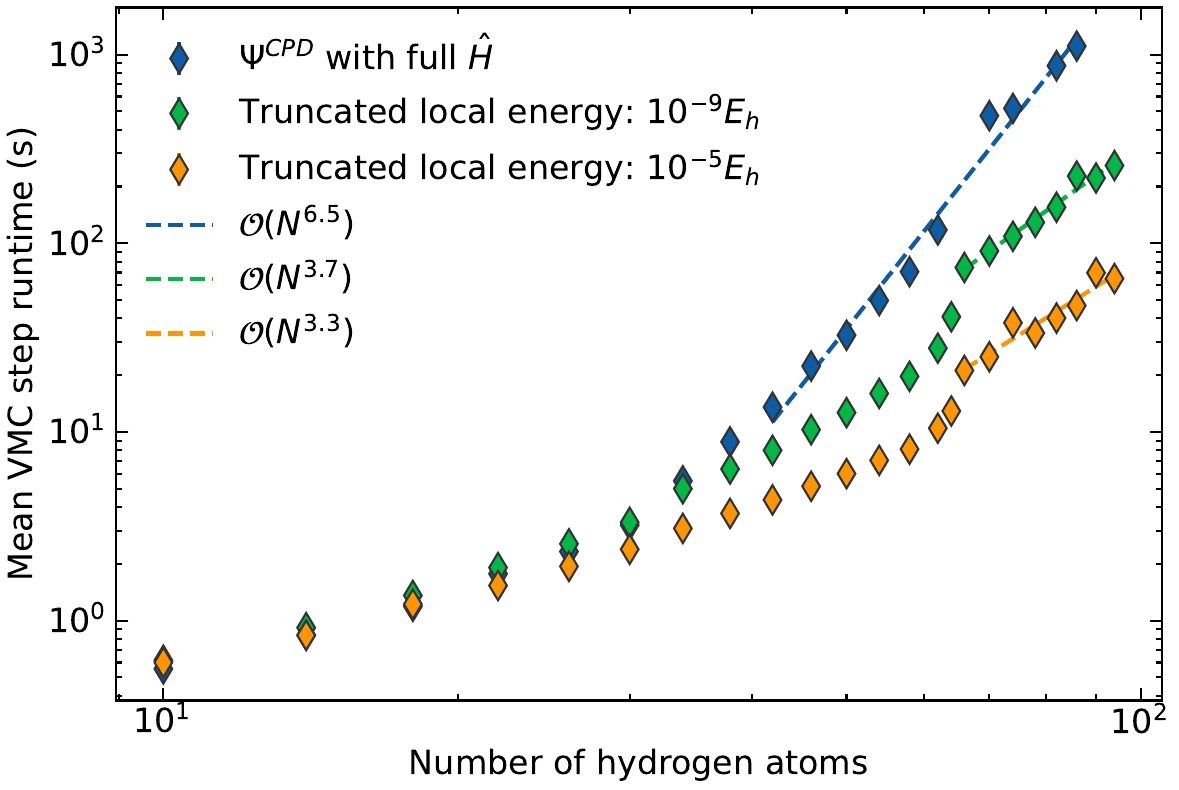}
    \caption{\vtwo{\textbf{Scalability of the CPD backflow ansatz.}}
    Mean VMC step runtime for the CPD backflow ansatz as a function of the number of atoms in a hydrogen chain with fixed inter-atomic distances ($1.68$ \AA) in a STO-6G basis.
    The support dimension is $M=1$ and the sample size is $N_S=128$.
    The blue points are obtained with the full Hamiltonian, while the green and orange points are obtained with the local energy evaluated with a Hamiltonian pruned with a threshold of $10^{-9}$ and $10^{-5} E_h$.
    The dashed lines show the observed scaling for the largest system sizes.}
    \label{fig:scalability}
\end{figure}

We show this scaling explicitly in Fig.~\ref{fig:scalability}, where we measure the mean runtime for a full VMC parameter update step for the CPD backflow ansatz, including the Markov chain sampling of a fixed number of configurations, evaluation of the local energy, and the subsequent parameter update.
By increasing the number of hydrogen atoms in a chain with fixed equilibrium inter-atomic distances ($d=1.68$ \AA) in a STO-6G basis up to 90 atoms we can extract a realistic asymptotic scaling of the approach.
We set the support dimension of the ansatz to $M=1$ and choose a sample size of $N_S=128$, in order to fit the data in memory even for the largest system sizes.
For each system size, we let the VMC algorithm run for 50 iterations on a single Nvidia A100 GPU with 40GB of memory.
Extracting the scaling from the large system limit gives a scaling of $\mathcal{O}[N^{6.5}]$, which is only evident for this system when we reach $>40$ atoms.

As discussed in the context of the GPS model in Ref.~\cite{rathFrameworkEfficientInitio2023} and building on other works in this area \cite{mahajanEfficientLocalEnergy2020,weiReducedScalingHilbert2018,sabzevariImprovedSpeedScaling2018,hachmannMultireferenceCorrelationLong2006}, this scaling for {\em ab initio} systems can be reduced further by truncating the number of terms in the sum over connected configurations at each evaluation of the local energy.
This truncation is performed on the magnitude of the Hamiltonian matrix element connecting the configurations.
By presorting the electron repulsion integrals between the degrees of freedom, this truncation can be implemented without having to consider the entire set of Hamiltonian matrix elements for each evaluation of the local energy.
Formally, the exponentially decreasing overlap between the orbitals in the sampled space should reduce the number of connected determinants which contribute to the local energy asymptotically to $\mathcal{O}[N^2]$ rather than $\mathcal{O}[N^4]$ -- a scaling which then matches the scaling of the local energy evaluation in a first quantized perspective. This results in a practical scaling of the CPD ansatz of $\sim\mathcal{O}[N^{4-6}]$.
However, since this method comes with a certain overhead in terms of data structures, the lower-bounds on this scaling only materialize after a certain crossover system size, which depends on the system and on the truncation threshold on the Hamiltonian matrix elements. 

Figure~\ref{fig:scalability} also shows the analogous results including this energetic threshold of $10^{-5}$ and $10^{-9} E_h$, with this tighter threshold expected to incur negligible change in the sampled energy for a given state.
The results show that a practical crossover point, after which the pruning of Hamiltonian elements below a small threshold yields a speed-up, is reached already around system sizes as low as 20-30 electrons for this ansatz, noting that a one-dimensional chain is advantageous in terms of affecting an advantage from this approach.
However, for large system sizes the speed-up is more than an order of magnitude and provides the expected asymptotic quadratic improvement, giving an overall scaling of $\mathcal{O}[N^{3-4}]$ up to $\sim100$ electrons, a scaling competitive with hybrid Density Functional Theory (DFT) techniques.
Overall, the combination of this ansatz with the various scaling reductions outlined represent a real potential towards a second quantized, systematically improvable, VMC algorithm with a practical and competitive $\mathcal{O}(N^{4})$ scaling for medium to large {\em ab initio} systems.
Clearly, further developments for prefactor reductions are key to take advantage of this improved scaling and access these system sizes.

\section{Conclusions and outlook}

In this work, we introduce a general and simple ansatz suitable for {\em ab initio} fermions, based on a systematically improvable tensor rank decomposition of a general backflow form.
This systematically builds configuration-dependent orbitals of a single antisymmetric Slater determinant in second quantization, directly encoding non-trivial $N$-body electron-electron correlations with a parameter scaling of $\mathcal{O}[N^{2-3}]$.
We have shown that the ansatz can achieve competitive accuracy on small Fermionic systems, such as the Fermi-Hubbard model and the water molecule, and that it can be used to model larger strongly correlated lattices of {\em ab initio} hydrogen atoms with an accuracy comparable to state-of-the-art DMRG techniques.
Finally, we have discussed the scalability of the ansatz and shown that we can affect various reductions in a practical fashion to demonstrate $\mathcal{O}(N^{4})$ scaling on medium to large {\em ab initio} systems.

We are working on further improvements in the accuracy and efficiency of the ansatz, as well as taking advantage of the benefits of working in a second quantized formalism to integrate with multiscale methods and quantum embedding methodologies to provide a practical route in the modelling of truly extended systems within this CPD backflow framework~\cite{nusspickelEffectiveReconstructionExpectation2023,nusspickelEffectiveReconstructionExpectation2023}. \vtwo{These techniques could also be integrated within the `hidden fermion' model of correlated states as an alternative parameterization of the correlations~\cite{morenoFermionicWaveFunctions2022,liuUnifyingViewFermionic2024}. It is also natural to ask whether alternative tensor factorization techniques could be applied within the context of describing second-quantized backflow correlations. These could naturally be fitted into the framework described above, and will also be explored in the future.}

\vtwo{\section*{Data availability}
All the results presented in this work can be fully reproduced with the publicly available source code and input configurations.
Energies and model parameters are further available upon request to the corresponding author.}

\section*{Code Availability}
The code for this project is implemented as part of the publicly available \href{https://github.com/BoothGroup/GPSKet}{GPSKet} plugin for \href{https://github.com/netket/netket}{NetKet}~\cite{carleoNetKetMachineLearning2019,vicentiniNetKetMachineLearning2022}.
It is made available, together with configurations files to reproduce the figures in the paper, at \url{https://github.com/BoothGroup/GPSKet/tree/master/scripts/cpd-backflow}.

\begin{acknowledgments}
    The authors gratefully acknowledge support from the Air Force Office of Scientific Research under award number FA8655-22-1-7011. We are grateful to the UK Materials and Molecular Modelling Hub for computational resources, which is partially funded by EPSRC (EP/P020194/1 and EP/T022213/1). Furthermore, we acknowledge the use of the high performance computing environment CREATE at King’s College London~\cite{kingscollegelondone-researchteamKingComputationalResearch2022}.
\end{acknowledgments}

\bibliographystyle{unsrtnat}% Sort bibliography in order of appearance
%\bibliography{main}

\appendix

\section{Spin-spin Correlations}\label{app:spin-spin-correlations}
For a pair of atoms $a$ and $b$, we define the instantaneous spin-spin correlation function as
\begin{equation}\label{eq:spin-corr}
    \begin{split}
    \left\langle \hat{S}_{\vec{r}_a}^z\hat{S}_{\vec{r}_b}^z \right\rangle &= \frac{1}{4}\sum_{ijkl}P_{ij}^aP_{kl}^b\left(\Gamma_{ijkl}^{\alpha\alpha}-\Gamma_{ijkl}^{\alpha\beta}-\Gamma_{ijkl}^{\beta\alpha}+\Gamma_{ijkl}^{\beta\beta}\right) \\
    &+ \frac{1}{4}\sum_{ijk}P_{ik}^aP_{jk}^b\left(\gamma_{ij}^{\alpha}+\gamma_{ij}^{\beta}\right),
    \end{split}
\end{equation}
where $P^a$ and $P^b$ are projectors onto the orbital subspaces of atom $a$ and $b$, the indices $i,j,k,l$ are orbital labels, and $\gamma_{ij}^{\sigma}$ and $\Gamma_{ijkl}^{\sigma\tau}$ are spinned one- and two-body reduced density matrices (RDMs), respectively.

During the evaluation of the local energy, the spin-free one- and two-body RDMs, $\gamma_{ij}$ and $\Gamma_{ijkl}$, are computed, so that it pays off to express the previous equation in terms of these quantities.
This can be done using the following identities~\cite{boynAccurateSingletTriplet2021,yunFerromagneticDomainsLarge2023}:
\begin{align}
    \gamma_{ij}^{\alpha} &= \gamma_{ij}^{\beta} = \gamma_{ij} \\
    \Gamma_{ijkl}^{\alpha\alpha} &= \Gamma_{ijkl}^{\beta\beta} = \frac{1}{6}\left(\Gamma_{ijkl}-\Gamma_{kjil}\right)\\
    \Gamma_{ijkl}^{\alpha\beta} &= \Gamma_{ijkl}^{\beta\alpha} = \frac{1}{6}\left(2\Gamma_{ijkl}+\Gamma_{kjil}\right).
\end{align}
Then, Eq.~\ref{eq:spin-corr} can be written as
\begin{equation}
    \begin{split}
    \left\langle \hat{S}_{\vec{r}_a}^z\hat{S}_{\vec{r}_b}^z \right\rangle &= -\frac{1}{2}\sum_{ijkl}P_{ij}^aP_{kl}^b\left(\frac{\Gamma_{ijkl}}{6}+\frac{\Gamma_{kjil}}{3}\right) \\
    &+ \frac{1}{4}\sum_{ijk}P_{ik}^aP_{jk}^b\left(\gamma_{ij}\right).
    \end{split}
\end{equation}

Since the $6\times 6$ hydrogen lattice results of Section \ref{sec:hydrogen} are obtained in a minimal basis of localized orthogonal orbitals, $P^a$ projects exactly onto one local orbital used in the sampling basis of the VMC and can thus be written as
\begin{equation}
    P_{ij}^a =
    \begin{cases}
        1 & \text{if } i=a\text{ and } j=a,\\
        0 & \text{otherwise}.
    \end{cases}
\end{equation}

For a given a variational state $\ket{\psi}$, the one- and two-body RDMs are estimated via Monte Carlo estimation of local operators
\begin{align}
    \gamma_{ij}^{loc} &= \frac{\braket{\mathbf{n}|\hat{c}_{i}^{\dagger}\hat{c}_{j}|\psi}}{\braket{\mathbf{n}|\psi}}, \\
    \Gamma_{ijkl}^{loc} &= \frac{\braket{\mathbf{n}|\hat{c}_{i}^{\dagger}\hat{c}_{j}^{\dagger}\hat{c}_{k}\hat{c}_{l}|\psi}}{\braket{\mathbf{n}|\psi}},
\end{align}
Evaluating these local operators involves finding all the connected Fock configurations $\mathbf{n}'$ for a given $\mathbf{n}$ sampled from $\ket{\psi}$ according to the corresponding Born distribution $|\psi(\mathbf{n})|^2$.
In the case of the one-body RDM, connected configurations are obtained by moving one electron from orbital $j$ to orbital $i$, while for the two-body RDM, they are obtained by a two-electron move.
The Fermionic commutation relations then imply that a sign is picked up for every electron that is passed during the move, within the pre-established orbital ordering.
This can be accounted for by computing a parity factor $P_{\mathbf{n}\mathbf{n}'}$ for each pair of configurations $\mathbf{n}$ and $\mathbf{n}'$.

\end{document}